\documentclass[twocolumn,english,aps,superscriptaddress,showpacs,prb,floatfix,longbibliography]{revtex4-1}
\usepackage[T1]{fontenc}
\usepackage[latin9]{inputenc}
\usepackage{color}
\usepackage{babel}
\usepackage{amsmath}
\usepackage{amssymb}
\usepackage{graphicx}
\usepackage{esint}
\usepackage[unicode=true,pdfusetitle,
 bookmarks=true,bookmarksnumbered=false,bookmarksopen=false,
 breaklinks=true,pdfborder={0 0 0},pdfborderstyle={},backref=false,colorlinks=true]
 {hyperref}
\hypersetup{
 linkcolor=blue,citecolor=blue,urlcolor=blue}

\makeatletter
\usepackage{pslatex}

\usepackage{color}
\usepackage{babel}

\makeatother

\begin{document}

\title{Emergent dimerization and localization in disordered quantum chains}

\author{André P. Vieira}

\affiliation{Instituto de Física, Universidade de São Paulo, C.P. 66318, São Paulo,
SP, 05508-090, Brazil}

\author{José A. Hoyos}

\affiliation{Instituto de Física de São Carlos, Universidade de São Paulo, C.P. 369,
São Carlos, SP, 13560-970, Brazil}

\date{\today}
\begin{abstract}
We uncover a novel mechanism for inducing a gapful phase in interacting
many-body quantum chains. The mechanism is nonperturbative, being
triggered only in the presence of both strong interactions and strong
aperiodic (disordered) modulation. In the context of the critical
antiferromagnetic spin-1/2 XXZ chain, we identify an emerging dimerization
which removes the system from criticality and stabilizes the novel
phase. This mechanism is shown to be quite general in strongly interacting
quantum chains in the presence of strongly modulated quasiperiodic
disorder which is, surprisingly, perturbatively irrelevant. Finally,
we also characterize the associated quantum phase transition via the
corresponding critical exponents and thermodynamic properties.
\end{abstract}
\maketitle

\section{Introduction\label{sec:intro}}

The presence of quenched disorder in noninteracting quantum systems
may lead to localization phenomena both in the case of random elements,
as in the Anderson model\citep{Anderson1958}, and of deterministic
quasiperiodic modulation, as in the Aubry--André model\citep{Aubry1980}.
Recently, in the context of many-body localization\citep{Basko2006,Nandkishore2015,Altman2015},
the interplay between interactions and deterministic disorder has
gained renewed interest both from the theoretical point of view\citep{Iyer2013,Nag2017,Khemani2017,Lee2017}
and from its experimental realization in ultracold atom systems\citep{Roati2008,Schreiber2015,Luschen2017,Bordia2017}. 

These studies usually deal with translational-symmetry breaking introduced
by an incommensurate potential. In contrast, here we consider the
effects of aperiodic modulation introduced in the exchange couplings.
We show that, for a certain class of coupling arrangements, the ground
state is delocalized for weak interactions even in the strong-modulation
limit, but sufficiently strong interactions induce a novel zero-temperature
transition to an emergent aperiodic dimer phase with localized low-energy
excitations. 

For concreteness, we focus on the spin-$\frac{1}{2}$ XXZ chain defined
by the Hamiltonian
\begin{equation}
H=\sum_{i=1}^{L-1}J_{i}\left(S_{i}^{x}S_{i+1}^{x}+S_{i}^{y}S_{i+1}^{y}+\Delta S_{i}^{z}S_{i+1}^{z}\right),\label{eq:xxzmodel}
\end{equation}
 in which $S_{i}^{x,y,z}$ are spin-$\frac{1}{2}$ operators and we
assume antiferromagnetic couplings $J_{i}>0$ with an easy-plane anisotropy
$-1/\sqrt{2}<\Delta\leq1$.  Via a Jordan-Wigner transformation, it
is well-known that (\ref{eq:xxzmodel}) also describes one-dimensional
spinless fermions with hopping amplitude $\propto J_{i}$ and interaction
strength $\propto J_{i}\Delta$. Thus, we will also refer to the anisotropy
parameter $\Delta$ as the interaction strength.

In the thermodynamic limit, the ground state of the uniform (clean)
system ($J_{i}\equiv J$) is critical and low-energy excitations are
described as a spin (Luttinger) liquid with a dynamical critical exponent
$z_{{\rm clean}}=1$. It is perturbatively unstable against dimerization
(i.e, alternating couplings $J_{i}\equiv\left[1+\frac{1}{2}(-1)^{i}\delta\right]J$,
with a dimerization strength $\delta$), which produces an energy
gap $\Delta E\sim\left|\delta\right|$ above the ground state and
a finite correlation length diverging as $\xi\sim\left|\delta\right|^{-\nu}$
for $\delta\rightarrow0$ with a critical exponent\citep{luther75,kossow2012}
$\nu=2\left(\pi-\arccos\Delta\right)/(3\pi-4\arccos\Delta)$.

The clean critical system is also perturbatively unstable against
random disorder (i.e, couplings $J_{i}$ independently chosen from
a probability distribution with a nonzero width $\delta J$), as dictated
by the Harris criterion\citep{harris74,doty92}. However, there is
no energy gap and the Luttinger liquid is replaced by a random-singlet
spin liquid whose low-energy physics is governed by a critical infinite-randomness
fixed point with an infinite dynamical critical exponent\citep{fisher94,hoyos07}.
Introducing correlations between the random couplings can either slightly
change the critical behavior of the infinite-randomness fixed point\citep{rieger-correlated}
or stabilize a line of finite-disorder critical points along which
the dynamical exponent remains finite but larger than one\citep{hoyos2011,getelina-etal-prb16}.

The effects of deterministic disorder are expected to be similar.
Indeed, for perturbatively relevant geometric fluctuations, the ground
state of the clean system is replaced by a critical self-similar version
of a random-singlet state with an infinite dynamical exponent, just
as for uncorrelated random disorder\citep{vieira05a,vieira05b}. For
marginally relevant geometric fluctuations, on the other hand, the
dynamical exponent remains finite but larger than one, just as for
the line of finite-disorder fixed points appearing in correlated random
aperiodicity. 

\begin{figure}[b]
\begin{centering}
\includegraphics[clip,width=0.75\columnwidth]{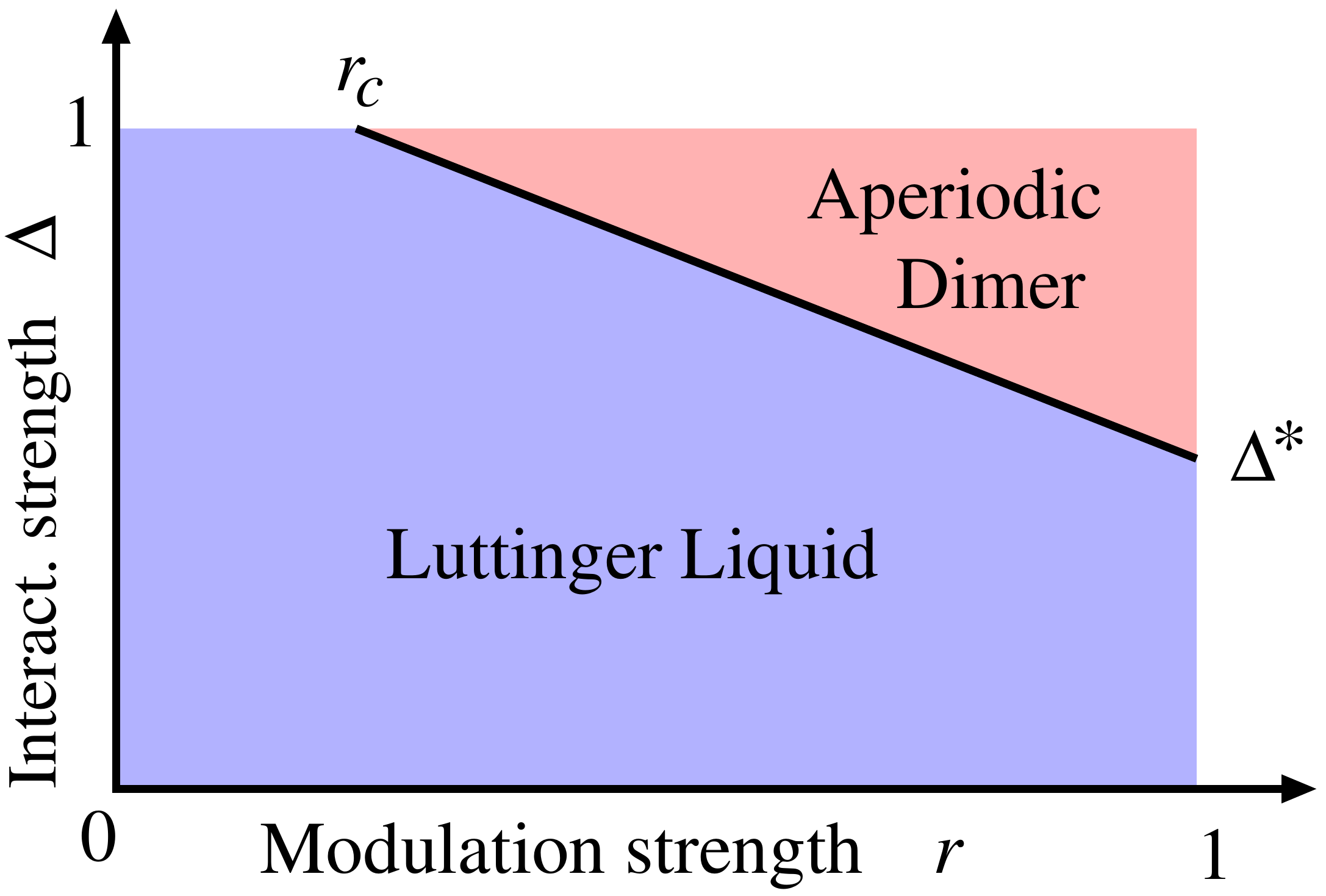}
\par\end{centering}
\caption{Schematic phase diagram of the XXZ spin-$\frac{1}{2}$ chain Eq.~(\ref{eq:xxzmodel})
in the presence of perturbatively irrelevant deterministic aperiodicity.
The anisotropy parameter $\Delta$ parameterizes the strength of the
interactions while the modulation $r$ parameterizes the strength
of the aperiodicity. For the aperiodic sequence defined in Eq.~(\ref{eq:seqomeganeg3}),
$r_{c}\approx0.13$ and $\Delta^{*}\approx0.69$.\label{fig:PD}}
\end{figure}

However, the case of perturbatively irrelevant deterministic disorder
has not been previously studied in detail. Evidently, for weak modulation
$r$ of the aperiodic couplings the system still corresponds to a
critical Luttinger liquid. In this paper we show that, surprisingly,
increasing $r$ beyond the perturbative limit ($r>r_{c}$) induces
the opening of an energy gap in the spectrum, as depicted in Fig.~\ref{fig:PD}.
To the best of our knowledge, this is the first verification that
an aperiodic perturbation (random or deterministic) induces such an
effect in a critical system. Furthermore, this effect is only possible
in the presence of sufficiently strong interactions (anisotropy parameter
$\Delta>\Delta^{*}$). Finally, we show that this gap is related to
an emergent dimerization of effective couplings in the low-energy
limit, characterizing an aperiodic dimer phase, with localized low-energy
excitations.

This paper is organized as follows. In Sec.~\ref{sec:2}, we introduce
the bond sequence at which we focus, representing the class of aperiodic
sequences defined in App.~\ref{sec:apA}, and discuss its perturbative
effects on the ground state of the quantum XXZ chain. In Sec.~\ref{sec:3},
we investigate the opposite limit of strong modulation. The results
of numerical calculations confirming the predictions in both limits
are reported in Sec.~\ref{sec:4}. Finally, Sec.~\ref{sec:5} presents
a discussion of our results, while some technical details are relegated
to Apps.~\ref{sec:apB} and \ref{sec:apC}.

\begin{figure*}[!t]
\begin{centering}
\includegraphics[clip,width=0.95\textwidth]{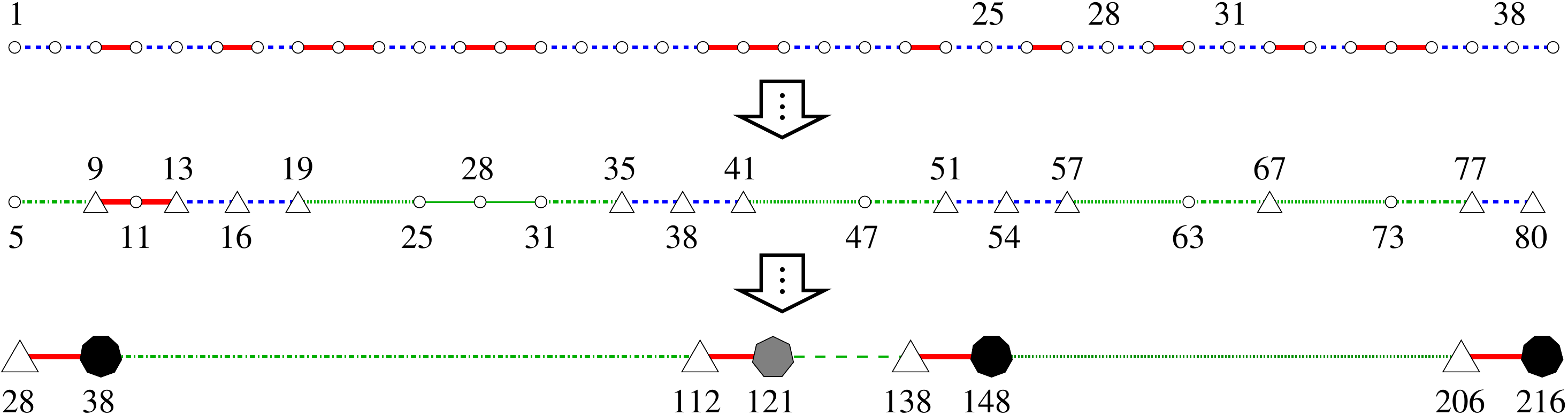}
\par\end{centering}
\caption{\label{fig:dimerization} The upper lattice shows the leftmost portion
of the aperiodic sequence of bonds in Eq.~(\ref{eq:seqomeganeg3}).
Dashed blue lines represent weak ($J^{(a)}$) couplings and solid
red lines represent strong ($J^{(b)}$) couplings. The middle lattice
is obtained after the first two renormalization steps (consisting
in decimating the strong bonds $J^{(b)}$ and all remaining weak bonds
$J^{\left(a\right)}$). The lower lattice shows the effective chain
produced in the latest stages of the SDRG method in the Heisenberg
limit $\Delta=1$. Notice the alternating pattern of strong (red/short/thick/solid)
and weak (green/long/thin/broken) effective couplings, revealing an
emergent dimerization. Circles and polygons represent real spins and
low-energy effective spin-1/2 degrees of freedom (effective spins),
respectively. In the latter case, the number of sides (3, 7, or 9)
is the number of real spins within. The numbers indicate the original
position of the central real spin.}
\end{figure*}

\section{Deterministic aperiodic sequences and their perturbative relevance\label{sec:2}}

Let us start by defining the main bond sequence $\{J_{i}\}$ investigated
in this work. Consider the following substitution rule for letter
pairs: 
\begin{equation}
\left\{ \begin{array}{ccc}
aa & \rightarrow & aa\,ba\,ab\,ab\,ba\\
ab & \rightarrow & aa\,ba\,ab\\
ba & \rightarrow & ab\,ba\,aa\,ab\,ba
\end{array}\right..\label{eq:seqomeganeg3}
\end{equation}
 Iterating this rule, starting from a single pair $aa$, we obtain
an aperiodic sequence of letters $a$ and $b$ which we associate,
respectively, with different bond values $J^{(a)}$ and $J^{(b)}$
of our aperiodic XXZ chain. The modulation of the aperiodic couplings
is quantified by $r\equiv1-J^{(a)}/J^{(b)}$. As detailed in App.~\ref{sec:apA},
the sequence in Eq.~(\ref{eq:seqomeganeg3}) represents a large family
of sequences exhibiting the same qualitative behavior, and was selected
on the basis of convenience for numerical calculations, since it gives
rise to a relatively large energy gap to the lowest excited states.

We emphasize the fact that there is no average dimerization induced
in the bonds $J_{i}$ by the substitution rule (\ref{eq:seqomeganeg3}),
the average couplings being the same at odd and even positions along
the chain. Therefore, no gap is expected for weak modulation $r$.

We studied the weak-modulation effects of couplings chosen from the
sequence in Eq.~(\ref{eq:seqomeganeg3}) by adapting the perturbative
renormalization-group (RG) method of Vidal, Mouhanna and Giamarchi\citep{vidal99,vidal01}
(see also Ref.~\onlinecite{hida01}) to the XXZ chain (\ref{eq:xxzmodel}).
In this approach, as described in Appendix \ref{sec:apB}, it is found
that the relevant effects of bond disorder on the corresponding low-energy
field theory describing the clean system are determined by the behavior
of the Fourier transform $\tilde{J}\left(Q\right)$ of the bond sequence
in the neighborhood of $Q=2k_{f}=\pi$, where $k_{f}=\pi/2$ gives
the location of the corresponding Fermi level. When the integrated
Fourier weight around $Q=\pi$ grows sufficiently slowly, the perturbative
RG approach predicts that weak modulation is irrelevant. For $0\leq\Delta\leq1$,
this is precisely the case of the whole family of aperiodic sequences
represented by the one in Eq.~(\ref{eq:seqomeganeg3}). This is consistent
with Luck's generalization\citep{luck93b} of the Harris criterion\citep{harris74}
for the perturbative relevance of aperiodicity on the critical behavior
of physical systems.

\section{The XXZ chain with an aperiodic bond distribution: strong modulation\label{sec:3}}

Having determined the perturbative irrelevance of our aperiodic system,
we now study its low-energy properties in the strong (non-perturbative)
modulation regime $r\approx1$ where an adaptation of the strong-disorder
real-space RG (SDRG) method\citep{ma79,bhatt-lee} for aperiodic XXZ
spin chains\citep{vieira05a,vieira05b} can be used. In this approach,
one identifies clusters of strongly coupled spins (the clusters connected
by solid red lines in Fig.\ref{fig:dimerization}). For $r\approx1$,
it is a good approximation to keep only the low-energy state of these
``molecules'' which is either a singlet (for $m$ even) or a doublet
(for $m$ odd), where $m$ is the number of spins in the molecule.
For a singlet, the molecule is simply removed from the effective chain
since its excitations are costly. In the case of a doublet, the molecule
is then replaced by a new effective spin-$\frac{1}{2}$ degree of
freedom (see the transition from the upper to the middle lattice in
Fig.\ref{fig:dimerization}). The new renormalized (and weaker) bonds
connecting the remaining spins in the lattice are obtained via perturbation
theory. Repeating this process, the energy scale is reduced and the
spatial distribution of couplings may reach a self-similar fixed point,
making it possible to write recursion relations for the effective
couplings and to obtain an approximate low-energy spectrum. (See Appendix
A of Ref.~\onlinecite{vieira05b} for details.) We mention that this
method was extended to higher spins\citep{casagrande2014}, to the
quantum Ising chain\citep{oliveira2012}, to the contact process \citep{barghathi2014},
and it can also be used to investigate entanglement properties\citep{igloi2007,juhasz2007}.

Let us now turn our attention back to the perturbatively irrelevant
sequence Eq.~(\ref{eq:seqomeganeg3}), to which we numerically apply
the SDRG method. Here no self-similar fixed-point exists. In the noninteracting
XX limit ($\Delta=0$), we find that, for any modulation strength
$0<r<1$, the effective couplings approach each other as the RG procedure
is iterated. In other words, the SDRG flows towards the clean fixed
point $r^{*}=0$. Therefore, we conclude that our aperiodicity is
irrelevant in both the weak and the strong modulation regimes, as
depicted in Fig.~\ref{fig:PD}.

In contrast, the SDRG flow completely changes its character for $\Delta>\Delta^{*}\approx0.69$.
As illustrated in Fig.~\ref{fig:dimerization} for $\Delta=1$, the
effective low-energy chain exhibits an \emph{emergent dimerization}
pattern alternating weak and strong effective couplings. Surprisingly,
all the strong couplings have the same magnitude, so that an energy
gap in the spectrum must exist above the ground state, and we find
it to scale as 
\begin{equation}
\Delta E/J^{(b)}\sim\left(J^{(a)}/J^{(b)}\right)^{2}=\left(1-r\right)^{2}.\label{eq:gap}
\end{equation}
On the other hand, the weak couplings follow a broad distribution
of lengths $l$ and strengths $J_{{\rm weak}}$, which are related
by 
\begin{equation}
J_{{\rm weak}}\sim\exp(-\mu\ln^{2}(l/l_{0})),\label{eq:JvsL}
\end{equation}
 the constants $\mu$ and $l_{0}$ depending only on the anisotropy
$\Delta$. For more details, see App. \ref{sec:apC}.

In the strong-modulation limit of the Heisenberg chain, the effective
Hamiltonian of a system with $\ell$ effective spins ($\ell$ even
for convenience) can be written as
\begin{equation}
\tilde{H}=\tilde{J}_{\text{strong}}\sum_{j=1}^{\ell/2}\vec{S}_{2j-1}\cdot\vec{S}_{2j}+\sum_{j=1}^{\ell/2-1}\tilde{J}_{j}\vec{S}_{2j}\cdot\vec{S}_{2j+1},
\end{equation}
 in which all strong effective couplings $\tilde{J}_{\text{strong}}$
have the same intensity, much larger than the intensities of the weak
effective couplings $\tilde{J}_{j}$, all of which can be calculated
from a numerical implementation of the SDRG scheme. 

If all $\tilde{J}_{j}$ were zero, the ground state could be written
as
\begin{equation}
\left|\Psi_{0}\right\rangle =\left|s\right\rangle _{1,2}\otimes\left|s\right\rangle _{3,4}\otimes\left|s\right\rangle _{5,6}\otimes\cdots\otimes\left|s\right\rangle _{\ell-1,\ell},
\end{equation}
 where $\left|s\right\rangle _{i,j}$ is a spin-$\frac{1}{2}$ singlet
between the effective spins $i$ and $j$. The perturbative effect
of the weak effective couplings on the ground state is to provide
a second-order correction to the ground-state energy. 

Again if all $\tilde{J}_{j}$ were zero, the $\frac{3}{2}\ell$ degenerate
lowest-energy excitations would correspond to states
\begin{equation}
\left|j_{S^{z}}\right\rangle =\left(\bigotimes_{i\neq j}\left|s\right\rangle _{2i-1,2i}\right)\otimes\left|t_{S^{z}}\right\rangle _{2j-1,2j},
\end{equation}
 with $\left|t_{S^{z}}\right\rangle _{i,j}$ denoting the triplet
state between the effective spins $i$ and $j$, and $S^{z}=-1,\ 0,\ 1$.
The degeneracy is then lifted when the weak couplings are turned on.
The first-order perturbative effective Hamiltonian for the lowest-energy
many-body band is a simple tight-binding chain with zero onsite potential
describing the hoppings of the ``triplons'' over the dimers, 
\begin{equation}
\tilde{H}_{\text{low}}=-\frac{1}{2}\sum_{j=1}^{\ell/2-1}\tilde{J_{j}}\left(\left|j_{S^{z}}\right\rangle \left\langle j+1_{S^{z}}\right|+\left|j+1_{S^{z}}\right\rangle \left\langle j_{S^{z}}\right|\right),\label{eq:TB=00003DH}
\end{equation}
 where an unimportant constant (the average gap to the ground state)
was neglected.

\begin{figure}
\begin{centering}
\includegraphics[width=0.99\columnwidth]{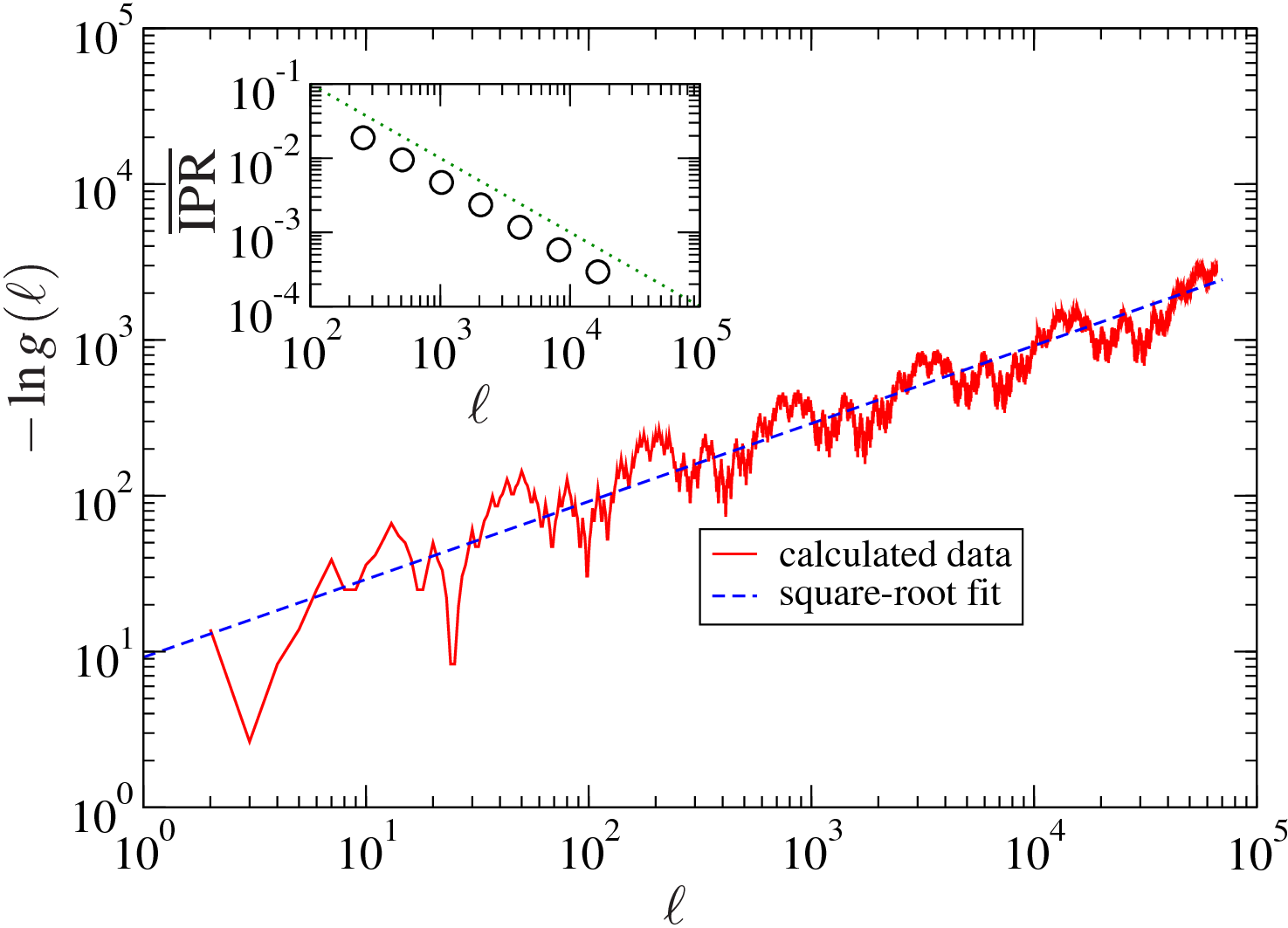}
\par\end{centering}
\caption{\label{fig:g_acopfracos}Main plot: Logarithm of the conductance as
a function of the number of effective spins $\ell$ of the effective
tight-binding Hamiltonian (\ref{eq:TB=00003DH}) describing the lowest-energy
many-body excitation band of the Heisenberg chain with couplings following
the sequence in Eq.~(\ref{eq:seqomeganeg3}) in the strong-modulation
limit. The blue dashed curve is a fit given by $-9.20\ell^{\psi}$,
with $\psi=\frac{1}{2}$. The chemical potential corresponds to the
band center. Inset: Average inverse participation as a function of
the effective system size. The green dotted line is proportional to
$1/\ell$.}
\end{figure}
We now ask whether the triplons are localized or not. From the set
$\left\{ \tilde{J}_{j}\right\} $ provided by the SDRG approach, and
following the analysis in Ref.~\onlinecite{Mard2014}, we study,
as a function of the effective system size $\ell$, the zero-temperature
conductance when the chemical potential corresponds to the center
of the first excited band, 
\begin{equation}
g\left(\ell\right)=\frac{4t_{\ell}^{2}}{\left(1-t_{\ell}^{2}\right)^{2}},
\end{equation}
where
\begin{equation}
t_{\ell}=-\frac{1}{2}\frac{\tilde{J}_{1}\tilde{J}_{3}\tilde{J}_{5}\cdots\tilde{J}_{\frac{1}{2}\ell-1}}{\tilde{J}_{2}\tilde{J}_{4}\tilde{J}_{6}\cdots\tilde{J}_{\frac{1}{2}\ell-2}}
\end{equation}
 is the system transmission coefficient (for convenience, we are assuming
that $\frac{1}{2}\ell$ is even). We choose this particular chemical
potential because it is expected to probe the least localized state
in the band \citep{eggarter78}. As shown in Fig.~\ref{fig:g_acopfracos},
$g(\ell)$ is compatible with the stretched-exponential scaling form
\begin{equation}
\ln g\left(\ell\right)\sim-\ell^{\psi},
\end{equation}
 with a tunneling exponent\citep{Mard2014} $\psi=\frac{1}{2}$. (The
superimposed log-periodic oscillations are a common feature of aperiodicity
generated by substitution rules.) As shown in Refs.~\onlinecite{vieira05a,vieira05b},
in the present context the tunneling exponent $\psi$ is related to
the pair wandering exponent $\omega_{\text{weak}}$ of the effective
weak couplings $\{\tilde{J}_{i}\}$ via $\psi=\omega_{\text{weak}}$.
We have explicitly verified that $\omega_{\text{weak}}=\frac{1}{2}$.
This is somewhat surprising. The effective aperiodic sequence of the
effective weak couplings emulate the effects of random aperiodicity,
characterized by $\omega_{\text{random}}=\frac{1}{2}$.

In addition, via exact diagonalization of the Hamiltonian (\ref{eq:TB=00003DH}),
we also computed the participation ratio 
\begin{equation}
p_{k}=\sum_{j}\left|\phi_{k,j}\right|^{4},
\end{equation}
 where $\phi_{k,j}$ is the corresponding wavefunction amplitude of
the $k$th eigenstate at ``site'' $j$. For an extended state, we
expect $p_{k}\sim1/\ell$, while for a localized state we should have
a $p_{k}$ of order unity. It is then convenient to calculate the
average inverse participation ratio,
\begin{equation}
\overline{\text{IPR}}=\frac{1}{\ell^{2}}\sum_{k}p_{k}^{-1}.
\end{equation}
If this quantity scales as $1/\ell$ for large $\ell$, the fraction
of extended states in the band is zero in the thermodynamic limit.
As shown in the inset of Fig.~\ref{fig:g_acopfracos}, this is precisely
what we obtain for the one-triplon band.

We then conclude that in the strong-modulation limit the lowest-energy
sector of the dimerized phase is localized. 

The SDRG results (\ref{eq:gap}) and (\ref{eq:JvsL}), being perturbative
in $J^{\left(a\right)}/J^{\left(b\right)}$, are not expected to hold
in the weak-modulation limit $r\ll1$. As shown in Eq.~(\ref{eq:gap}),
the SDRG scheme predicts a monotonically decreasing energy gap $\Delta E$
as a function of the modulation strength $r$. However, a nonmonotonic
behavior is expected since the system is critical for $r\ll1$. In
the simplest scenario of a single critical point, increasing the modulation
starting from the clean system ($r=0$) and $\Delta>\Delta^{*}$ we
expect a gap opening at $r=r_{c}>0$, then reaching a maximum, and
finally vanishing as in Eq.~(\ref{eq:gap}).

\section{Unbiased numerical results\label{sec:4}}

In order to check the predictions for weak versus strong modulation,
we resort to unbiased numerical methods, focusing on the aperiodic
sequence in Eq.~(\ref{eq:seqomeganeg3}). We measure energies in
units of $J^{(b)}$ for various modulation strengths $r=1-J^{(a)}/J^{(b)}$,
and taking $0<J^{(a)}<J^{(b)}$. 

Using the Jordan-Wigner fermionization method\citep{lieb61}, we studied
the XX chain ($\Delta=0$) through exact numerical diagonalization
of very large system sizes ($L\sim10^{6}$) and fully confirmed the
predictions of both the perturbative and the SDRG methods that the
clean critical system is robust against aperiodicity for any modulation
strength. This indicates that there is no single-particle localization
at low energies.

\begin{figure}[b]
\begin{centering}
\includegraphics[width=0.95\columnwidth]{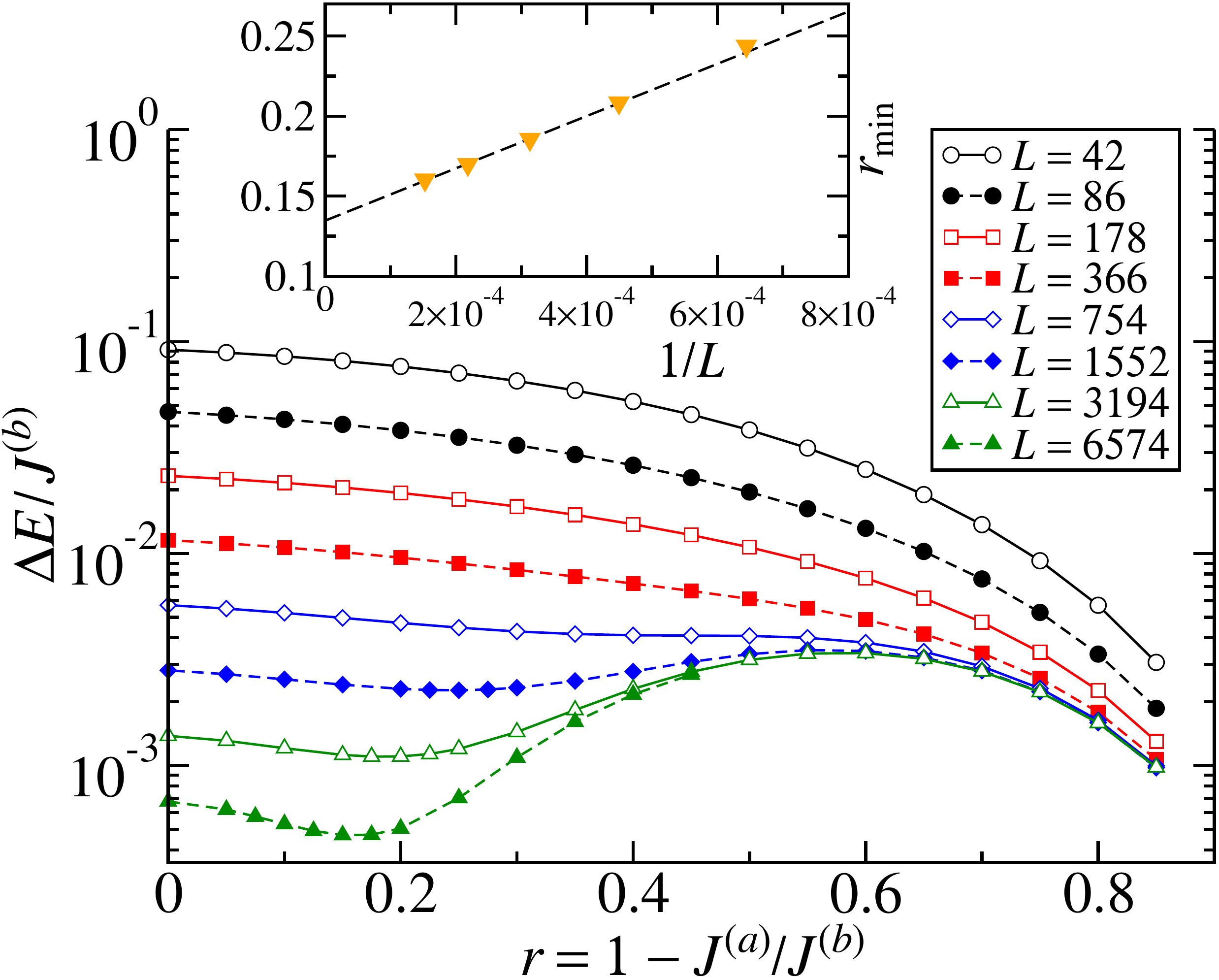}
\par\end{centering}
\caption{\label{fig:gapvsr_omeganeg3}Linear-log plot the energy gap as a function
of the coupling modulation $r$ for the Heisenberg chain with couplings
following the sequence in Eq.~(\ref{eq:seqomeganeg3}) for various
chain lengths $L$. The inset shows the position of the relative minimum
in the curves for large $L$, using also the intermediate values $L=2226$
and $L=4582$ (not shown for the sake of clarity).}
\end{figure}

In order to investigate the Heisenberg chain ($\Delta=1$), we performed
numerical calculations using the quantum Monte Carlo (QMC) and the
density-matrix renormalization group (DMRG) algorithms from the ALPS
project\citep{ALPS,bauer2011}.

We employed the DMRG method for calculating the energy gap $\Delta E$
defined as the energy difference between the ground state (with total
spin $S_{T}=0$) and the first excited state ($S_{T}=1$), using even
lattice sizes ranging from $L=42$ to $6\,574$. Except for the largest
chain size, we used up to $50$ warm-up states to grow the DMRG blocks,
keeping a maximum of up to $500$ SU(2) states during the (up to 20)
sweeps. For $N=6\,574$, we used up to $100$ warm-up sates and $1\,000$
SU(2) states during $30$ sweeps. The modulation strength was varied
starting from $r=0$ to $0.85$ and we increased the above simulational
parameters from their default values until the energies for each state
converged within a relative error below $10^{-8}$. For $r>0.85$
convergence could not be obtained with the maximum values of the above
parameters. For the largest system size studied ($N=6\,574$), despite
the higher number of states kept, convergence of the gaps was still
poorer than for smaller sizes, and we estimate a higher relative error
around $10^{-4}$. Figure \ref{fig:gapvsr_omeganeg3} shows the results
of these calculations for various chain lengths. For $r\approx0$
the finite-size gaps scale as $L^{-z}$ with a (clean) dynamical exponent
$z=z_{{\rm clean}}=1$, whereas for larger $r$ they converge to a
finite value exhibiting a maximum $\approx3\times10^{-3}J^{(b)}$
at $r\approx0.6$. For large $L$, local minima are visible near $r\approx0.2$.
Their precise positions are obtained from quadratic fits and plotted
as a function of $1/L$ in the inset. From a linear extrapolation,
we conclude that the minimum occurs at $r_{c}\approx0.135$ for $L\rightarrow\infty$,
therefore supporting the existence of a finite range $0\leq r\leq r_{c}$
for which the system is gapless in the thermodynamic limit. This is
in agreement with the simplest scenario of a single critical point
and with our perturbative RG predictions. The appearance of a gap
only for sufficiently strong modulation is consistent with the emergent
dimerization scenario predicted by the SDRG method. A sketch of a
generic phase diagram is given in Fig.~\ref{fig:PD}. 

\begin{figure}
\begin{centering}
\includegraphics[width=0.99\columnwidth]{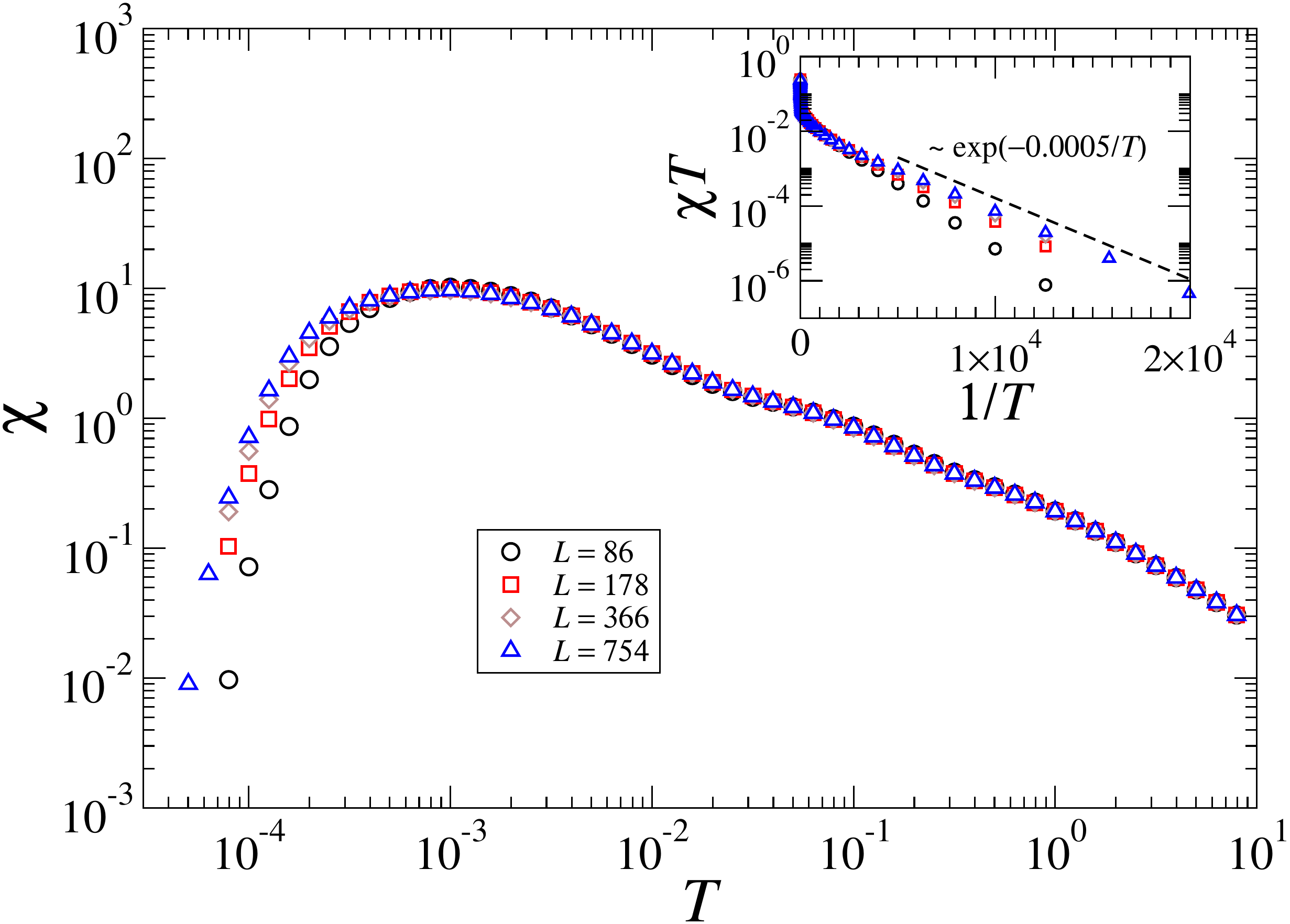}
\par\end{centering}
\caption{\label{fig:susc}Susceptibility $\chi$ as a function of temperature
$T$ for the Heisenberg chain with couplings following the sequence
in Eq. (\ref{eq:seqomeganeg3}), for a coupling ratio $J^{\left(a\right)}/J^{\left(b\right)}=1/10$.
The results were obtained by using the SSE QMC algorithm, with open
chains containing $N\in\left\{ 86,178,366,754\right\} $ spins. The
inset shows that the product $\chi T$ follows $\exp\left(-\Delta E/T\right)$,
with a size-dependent energy gap $\Delta E$ which approaches $\approx5\times10^{-4}J_{b}$
as $N\rightarrow\infty$. Temperature is measured in units of $J^{\left(b\right)}/k_{B}$.
The oscillations in $d\chi/dT$ for temperatures between $T\simeq10$
and $T\simeq10^{-3}$ reflect the energy scales associated with the
formation of ``spin molecules'', as predicted by the SDRG scheme (see
main text).}
\end{figure}

The existence of a gap for strong modulation in the Heisenberg limit
is also confirmed by QMC calculations based on the stochastic series
expansion (SSE) algorithm\citep{Alet05,ALPS} with up to $2\times10^{5}$
thermalization steps and $10^{6}$ sweeps. Figure \ref{fig:susc}
shows the results of QMC calculations of the magnetic susceptibility
$\chi$ for a coupling ratio $J^{\left(a\right)}/J^{\left(b\right)}=1/10$,
with open chains containing from $86$ to $754$ spins. At low temperatures,
the results conform to the expected behavior
\begin{equation}
\chi\left(T\right)\sim\frac{e^{-\Delta E/T}}{T}
\end{equation}
 in the presence of an energy gap $\Delta E$. As shown in the inset,
the estimated value of the gap ($\approx5\times10^{-4}J^{(b)}$) is
compatible with those provided by the DMRG calculations. We also performed
QMC calculations for $J^{\left(a\right)}/J^{\left(b\right)}=4/10$
and $J^{\left(a\right)}/J^{\left(b\right)}=9/10$ (not shown), again
obtaining energy gaps compatible with those provided by DMRG.

\begin{figure}[t]
\begin{centering}
\includegraphics[width=0.95\columnwidth]{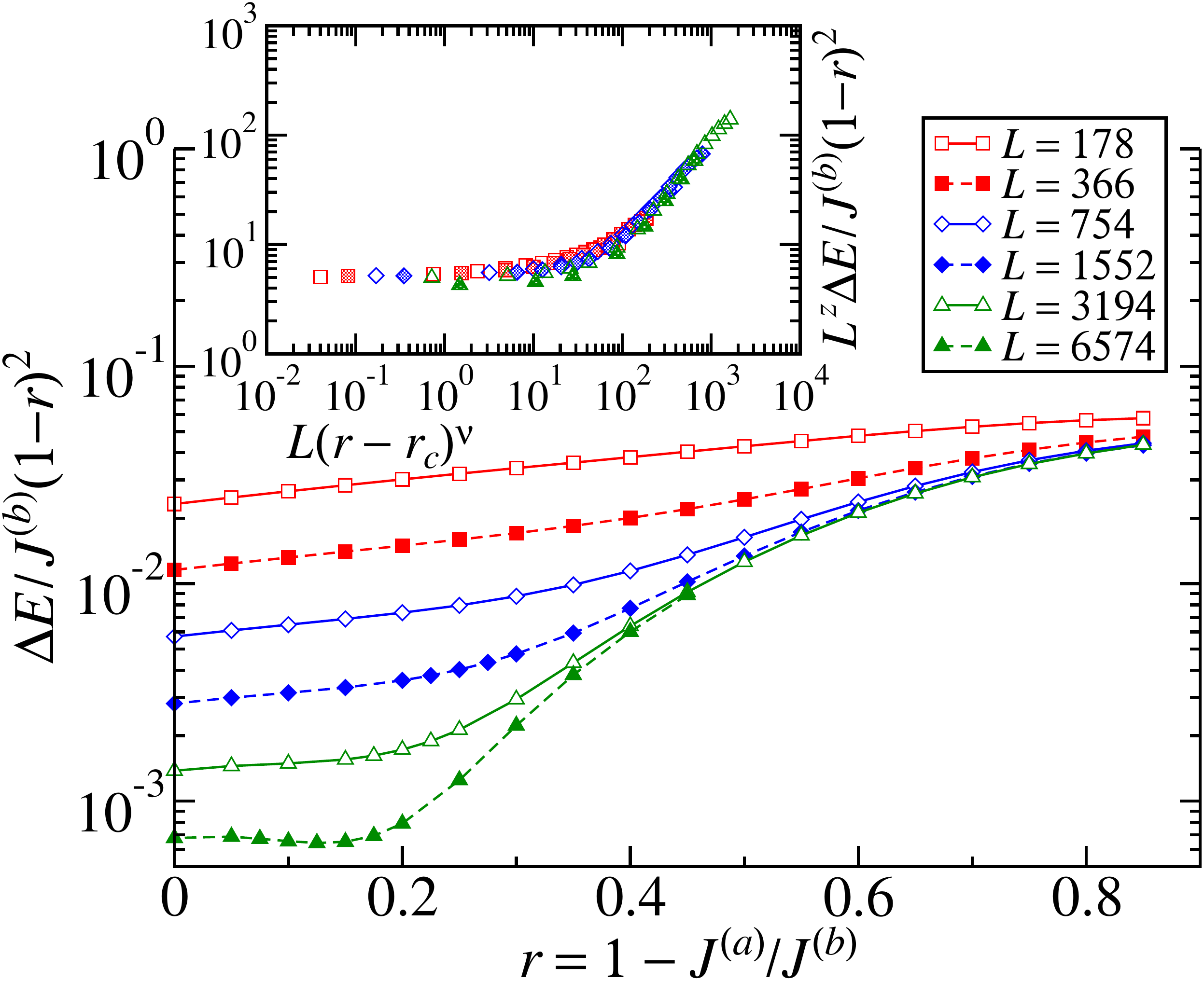}
\par\end{centering}
\caption{\label{fig:gapvsr_omeganeg3_sdrg} Same data as in Fig.~\ref{fig:gapvsr_omeganeg3}
with the energy gap rescaled by $\left(1-r\right)^{2}$. The inset
shows the data collapse obtained by the finite-size scaling hypothesis
in Eq.~(\ref{eq:gapfss}) using $r>r_{c}=0.135$, $z=1$ and $\nu=2$.}
\end{figure}

Rescaling the finite-size gaps by the asymptotic dependence $\sim\left(1-r\right)^{2}$
in Eq.~(\ref{eq:gap}), a monotonic behavior of the rescaled gaps
with $L$ and $r>r_{c}$ becomes manifest, as shown in Fig.~\ref{fig:gapvsr_omeganeg3_sdrg}.
It then suggests that a data collapse with a finite-size scaling hypothesis
may be possible. For $r>r_{c}$, we expect that in the thermodynamic
limit the gap scales as $\Delta E_{\infty}\sim\xi^{-z}\sim\left(r-r_{c}\right)^{z\nu}$,
in which $\xi$ is the correlation length, while $z$ and $\nu$ are
critical exponents. This gives rise to a finite-size scaling hypothesis
\begin{equation}
\Delta E_{N}=\xi^{-z}F(\xi/L)=L{}^{-z}{\cal F}\left(L\left(r-r_{c}\right)^{\nu}\right),\label{eq:gapfss}
\end{equation}
with scaling functions $F\left(x\right)$ and ${\cal F}\left(x\right)=x^{z}F\left(1/x\right)$. 

The plots in the inset of Fig.~\ref{fig:gapvsr_omeganeg3_sdrg},
obtained with $r_{c}=0.135$, $z=z_{{\rm clean}}=1$ and $\nu=2$,
show that our DMRG data are compatible with Eq.~(\ref{eq:gapfss}).
{[}Close to the critical point the rescaling of the data by $\left(1-r\right)^{2}$
becomes irrelevant.{]} This strongly suggests that a true phase transition
takes place and that the system is indeed gapless for $r<r_{c}$,
in agreement with the perturbative RG prediction. 

\section{Conclusions\label{sec:5}}

We showed that the interplay between strong modulation and interactions
induces a transition to a gapped phase in a broad class of deterministic
disordered (aperiodic) spin-1/2 chains. In this phase we identify
a surprisingly emergent dimerization of the effective low-energy chain,
which is quite distinct from any other known gap-inducing mechanism,
such as the explicit introduction of dimerization or the spin-Peierls
(Majumdar-Ghosh) mechanism related to a spontaneous breaking of translational
symmetry via spin-phonon (sufficiently strong frustrating next-nearest-neighbor
or, for $S>1/2$, biquadratic) interactions. 

Deep inside the aperiodic dimer phase, we showed that the first excited
many-body band (corresponding to one-triplon excitations) is entirely
localized. Whether these lowest-energy quasiparticle excitations remain
localized throughout the entire dimerized phase is a topic left for
future research. Another question we leave for future investigation
is whether the higher-energy bands also harbor localized states.

Having characterized our zero-temperature phase transition, and in
view of the recent evidence that localized ground states correspond
to many-body localized excited eigenstates of related Hamiltonians\citep{Dupont2018},
we hope our model may be useful to shed light on the nature of the
many-body localization transition, which remains largely unclear \citep{Nandkishore2015,Altman2015}. 

Our results also apply to interacting fermionic models which are equivalent
to the quantum spin chains explicitly discussed here, and in principle
could be put to experimental test in the context of cold-atom systems.
Along the lines discussed in Ref.~\onlinecite{Duan2003}, that would
involve trapping fermionic atoms in optical lattices. The antiferromagnetic
interactions between effective spin degrees of freedom would be related
to the atomic tunneling rates between neighboring extrema of the light
patterns, and these rates could be made aperiodic by employing several
laser sources\citep{Jagannathan2014}, possibly in combination with
a cut-and-project construction\citep{Singh2015}. The local intensities
at the potential extrema must also be controlled, which could be arranged
by employing a digital mirror device\citep{Choi2016}.

Finally, we point out that, in the fermion context, the transition
we identified is a metal-insulator transition very distinct from the
conventional cases of the Mott and the Anderson transitions. It is
driven by both strong interactions and disorder modulation, yielding
a fundamentally different insulating phase which has no charge order
and exhibits a spectral gap, the first excited band being localized.
\begin{acknowledgments}
This work was supported by the Brazilian agencies FAPESP and CNPq.
JAH also acknowledges the hospitality of the Aspen Center for Physics,
and the financial support of the NSF and the Simons Foundation. We
thank Thomas Vojta for useful discussions.
\end{acknowledgments}

\appendix

\section{The Harris--Luck criterion, aperiodic sequences and geometric fluctuations\label{sec:apA}}

In order to determine the stability of a clean critical system against
the perturbative effects of aperiodicity ($\left|r\right|\ll1$),
Luck\citep{luck93b} generalized the Harris criterion\citep{harris74}
for the case of deterministic disorder. In the present context, one
then quantifies the geometric fluctuations of nonoverlapping letter
pairs via the wandering exponent $\omega<1$ defined by 
\begin{equation}
G\left(N\right)\equiv\left|N^{\left(aa\right)}-p_{aa}N\right|\sim N^{\omega},\label{eq:GN}
\end{equation}
in which $N^{\left(aa\right)}$ denotes the number of $aa$ pairs
in the sequence built from cutting the infinite sequence at the $N$th
pair, and $p_{aa}$ is the expected fraction of $aa$ pairs in the
$N\rightarrow\infty$ limit. Once $\omega$ is determined, the Harris--Luck
criterion states that, necessarily, for a clean critical point to
be stable against aperiodic weak modulation, the wandering exponent
must fulfill 
\begin{equation}
\omega<\omega_{c}=\max\left\{ 0,1-(d\nu)^{-1}\right\} ,\label{eq:Harris-Luck}
\end{equation}
 where $d$ is the number of spatial dimensions in which the system
is disordered and $\nu$ is the correlation length critical exponent
of the clean theory. As a self-consistent criterion for the stability
of the clean fixed point, upon its violation the Harris--Luck criterion
does not tells us what is the low-energy physics replacing that of
the clean system. We mention that all cases previously studied indicate
that the system remains critical but with a larger dynamical exponent\citep{vieira05a,vieira05b}.

When fulfilled ($\omega<\omega_{c}$), the Harris--Luck criterion
suggests that the corresponding aperiodic sequence is an irrelevant
perturbation. Finally, for $\omega=\omega_{c}$ the perturbation is
marginal and thus nonuniversal effects may be expected. In this case,
a less general approach (as discussed later) is thus required for
determining the precise fate of the clean critical point.

We would like to stress the distinction between the strength of the
geometric fluctuations, gauged by the wandering exponent $\omega$,
and the strength of the aperiodic modulation $r=1-J^{(a)}/J^{(b)}$:
for a given $\omega$ (i.e., a given substitution rule) we can tune
the system from the clean limit ($r=0$) to the strong-modulation
regime ($r\rightarrow1$ or $r\rightarrow-\infty$). Consider, for
concreteness, the pair substitution rule giving rise to the Rudin--Shapiro
sequence, 
\begin{equation}
\left\{ \begin{array}{ccc}
aa & \rightarrow & aa\,ab\\
ab & \rightarrow & aa\,ba\\
ba & \rightarrow & bb\,ab\\
bb & \rightarrow & bb\,ba
\end{array}\right..\label{eq:seqrudin}
\end{equation}
 The geometric fluctuations of nonoverlapping letter pairs after $n$
iterations of the substitution rule are quantified by
\begin{equation}
G_{n}\equiv\left|N_{n}^{\left(aa\right)}-p_{aa}N_{n}\right|\sim N_{n}^{\omega_{\text{nat}}},\label{eq:Gn}
\end{equation}
 in which $N_{n}$ (called the \emph{natural} length of the sequence)
is the total number of letter pairs obtained after $n$ iterations
of the substitution rule, $N_{n}^{\left(\alpha\beta\right)}$ is the
corresponding number of $\alpha\beta$ pairs, $p_{\alpha\beta}$ is
the expected fraction of $\alpha\beta$ pairs in the $n\rightarrow\infty$
limit, and 
\begin{equation}
\omega_{\text{nat}}=\frac{\ln\left|\lambda_{2}\right|}{\ln\lambda_{1}}\label{eq:wnat}
\end{equation}
 is the natural wandering exponent\citep{luck93a}, $\lambda_{1}$
and $\lambda_{2}$ being, respectively, the two largest eigenvalues
(in absolute value) of the substitution matrix 
\begin{equation}
\mathbb{M}=\left(\begin{array}{cccc}
\#_{aa}(w_{aa}) & \#_{aa}(w_{ab}) & \#_{aa}(w_{ba}) & \#_{aa}(w_{bb})\\
\#_{ab}(w_{aa}) & \#_{ab}(w_{ab}) & \#_{ab}(w_{ba}) & \#_{ab}(w_{bb})\\
\#_{ba}(w_{aa}) & \#_{ba}(w_{ab}) & \#_{ba}(w_{ba}) & \#_{ba}(w_{bb})\\
\#_{bb}(w_{aa}) & \#_{bb}(w_{ab}) & \#_{bb}(w_{ba}) & \#_{bb}(w_{bb})
\end{array}\right),
\end{equation}
for which $\#_{\alpha\beta}\left(w_{\gamma\delta}\right)$ denotes
the number of $\alpha\beta$ pairs in the word associated with the
$\gamma\delta$ pair in the substitution rule. (Notice that $G_{n}$
could equally have been defined in terms of a different $\alpha\beta$
pair, which would not affect the value of $\omega_{\text{nat}}$.) 

It is important to notice the difference between the geometrical fluctuations
defined in Eqs.~(\ref{eq:GN}) and (\ref{eq:Gn}). Evidently, $G(N_{n})=G_{n}$.
Moreover, 
\begin{equation}
\omega\geq\omega_{\text{nat}}.\label{eq:w-wnat}
\end{equation}
 In order to illustrate the difference, we will compare $G(N)$ and
$G_{n}$ for different sequences. 

\begin{figure}
\begin{centering}
\includegraphics[width=0.99\columnwidth]{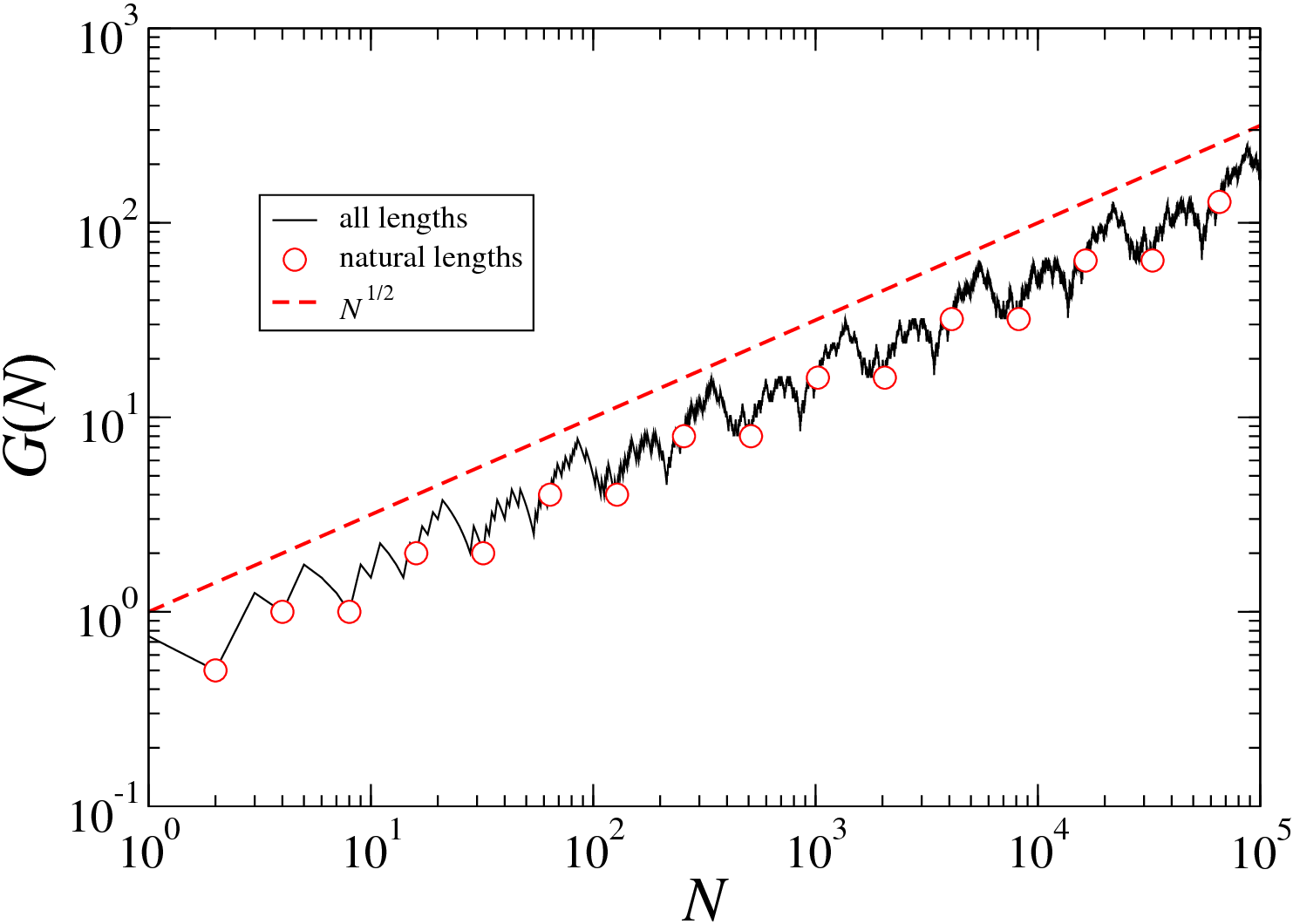}
\par\end{centering}
\caption{\label{fig:flutrudin}Comparison between geometric fluctuations induced
by the Rudin-Shapiro sequence Eq.~(\ref{eq:seqrudin}) as calculated
for all lengths (solid line) and only for the natural lengths of the
sequence (circles). The red dashed line is proportional to $N^{1/2}$.
Notice the existence of both stronger and weaker fluctuations in the
neighborhood of the natural lengths.}
\end{figure}

Let us start with the Rudin--Shapiro sequence (\ref{eq:seqrudin}),
for which $\omega_{\text{nat}}=\frac{1}{2}$. As plotted in Fig.~\ref{fig:flutrudin},
both geometric fluctuations $G(N)$ and $G_{n}$ scale as $N^{\omega}$
with $\omega=\omega_{\text{nat}}=\frac{1}{2}$. This equality between
$\omega_{\text{nat}}$ and $\omega$ can be verified for all the aperiodic
sequences generated by substitution rules with $\omega_{\text{nat}}>0$
investigated in Refs.~\onlinecite{vieira05a,vieira05b}.

We now turn our attention to the more involved case in which $\omega_{\text{nat}}\leq0$.
One paradigmatic example for $\omega_{\text{nat}}=0$ is the so-called
Fibonacci sequence defined (for letter pairs) by the substitution
rule
\begin{equation}
\left\{ \begin{array}{ccc}
aa & \rightarrow & ab\,aa\,ba\,ba\,ab\\
ab & \rightarrow & ab\,aa\,ba\,ba\\
ba & \rightarrow & ab\,aa\,ba\,ab
\end{array}\right..\label{eq:seqfib}
\end{equation}

\begin{figure}
\begin{centering}
\includegraphics[width=0.99\columnwidth]{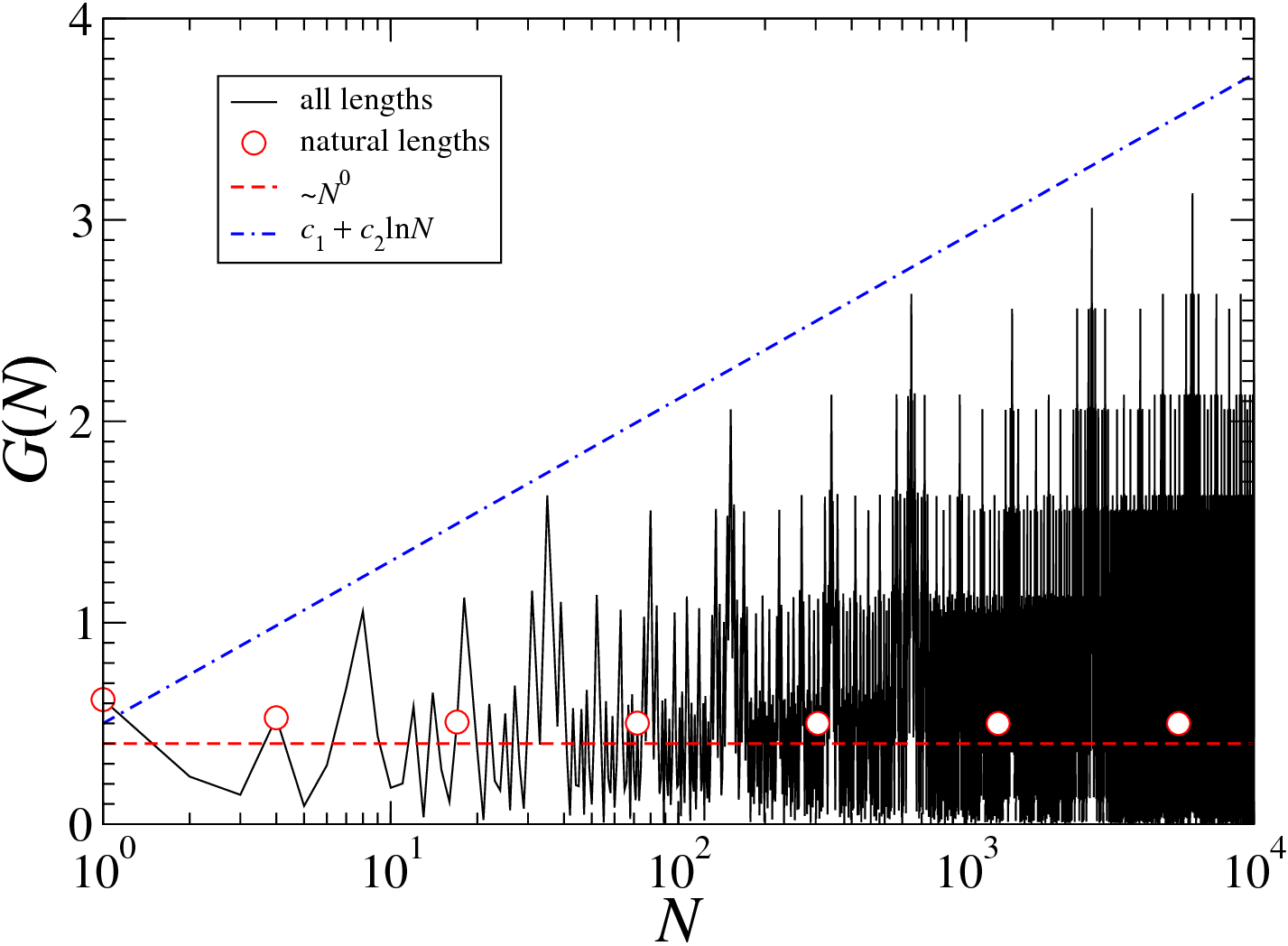}
\par\end{centering}
\caption{\label{fig:flutoparesfib}Comparison between geometric fluctuations
induced by the Fibonacci sequence (\ref{eq:seqfib}) as calculated
for all lengths (solid line) and only for the natural lengths of the
sequence (circles). The red dashed red line is proportional to $N^{0}$.
Notice that the stronger fluctuations scale at most logarithmically
with $N$ (blue dot-dashed line, with $c_{1}$ and $c_{2}$ constants
of order 1).}
\end{figure}

In this case, as shown in Fig.~\ref{fig:flutoparesfib}, the strong
fluctuations of $G(N)$ are unbounded but only grow logarithmically.
In this case, it is desirable to distinguish a logarithmic growth,
as for $G(N)$, from a constant, as for $G_{n}$. Here, we will simply
define the wandering exponent as $\omega=0^{+}$, which is still compatible
with Eq.~(\ref{eq:w-wnat}).

\begin{figure}
\begin{centering}
\includegraphics[width=0.99\columnwidth]{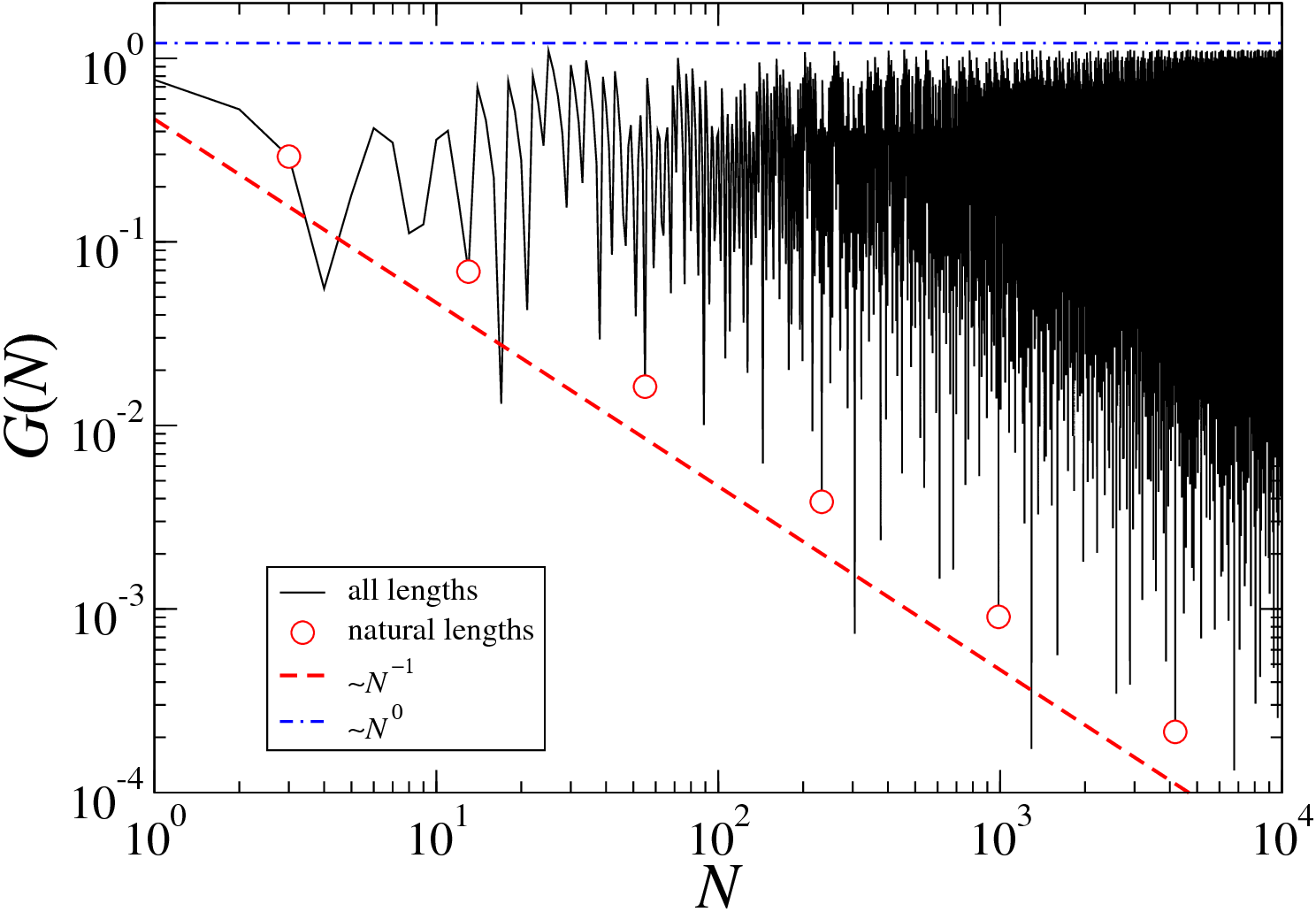}
\par\end{centering}
\caption{\label{fig:flutomeganeg3}Comparison between geometric fluctuations
induced by the sequence in Eq.~(\ref{eq:seqomeganeg3}) as calculated
for all lengths (solid line) and only for the natural lengths of the
sequence (circles). The red dashed curve is proportional to $N^{-1}$.
Notice that the stronger fluctuations scale as $N^{0}$ for large
$N$.}
\end{figure}

The sequences for which $\omega_{\text{nat}}<0$ (and finite), as
those of interest in this work, are said to exhibit the Pisot property~\citep{godreche92},
giving rise to \emph{bounded} fluctuations as $N\rightarrow\infty$.
Let us illustrate this case with the sequence defined by the substitution
rule in Eq.~(\ref{eq:seqomeganeg3}). The corresponding natural wandering
exponent $\omega_{\text{nat}}=-1$ obtained from Eq.~(\ref{eq:wnat})
is in agreement with the observed one shown in Fig.~\ref{fig:flutomeganeg3}.
In addition, notice that the strongest fluctuations (corresponding
to lengths other than the natural ones) are also bounded. For this
reason, we define the wandering exponent as $\omega=0^{-}$.

We would like to point out that, for our numerical analysis of the
sequence in Eq.~(2) of the main text, we used chains with lengths
not restricted to the natural ones. In other words, the striking features
we observed (as the gap behavior in Fig.~3 of the main text) are
not an artifact of choosing special chain lengths.

Finally, we mention the existence of many other sequences sharing
the same features of the sequence (\ref{eq:seqomeganeg3}), also yielding
the same emergent dimerization phenomena in the nonperturbative regime,
as reported in the main text. The sequences are such that $\omega=0^{-}>\omega_{\text{nat}}>-\infty$,
and do not induce an average dimerization. A simple way to construct
such a sequence is via small tweaks of the sequence in (\ref{eq:seqomeganeg3}),
as for example 
\begin{equation}
\left\{ \begin{array}{ccc}
aa & \rightarrow & aa\,ab\,ba\\
ab & \rightarrow & aa\,ba\,ab\,aa\\
ba & \rightarrow & aa\,ab\,ba
\end{array}\right.,\label{eq:seqomeganeg4}
\end{equation}
 which also yields $\omega_{\text{nat}}=-1$. Other examples are the
sequence 
\begin{equation}
\left\{ \begin{array}{ccc}
aa & \rightarrow & aa\,ab\,aa\,ba\\
ab & \rightarrow & ba\,ab\\
ba & \rightarrow & aa\,ba\,aa\,ab
\end{array}\right.,\label{eq:seqomeganeg5}
\end{equation}
 for which $\omega_{\text{nat}}=\frac{\ln\left(2-\sqrt{2}\right)}{\ln\left(2+\sqrt{2}\right)}\approx-0.44$,
and the sequence generated by the substitution rule
\begin{equation}
\left\{ \begin{array}{ccc}
aa & \rightarrow & aa\,ba\,ab\,ab\,ba\\
ab & \rightarrow & aa\,ab\,ba\\
ba & \rightarrow & aa\,ab\,aa\,ba\,ba\,ab
\end{array}\right.,\label{eq:omeganegseq}
\end{equation}
for which $\omega_{\text{nat}}=\frac{\ln\left|2-\sqrt{7}\right|}{\ln\left(2+\sqrt{7}\right)}\approx-0.28.$
Somewhat simpler sequences are given by the substitution rules
\begin{equation}
\left\{ \begin{array}{ccc}
aa & \rightarrow & aa\,ab\,ba\\
ab & \rightarrow & aa\,aa\\
ba & \rightarrow & ba\,ab\,aa
\end{array}\right.,\label{eq:seqsimple1}
\end{equation}
for which $\omega_{\text{nat}}=\frac{\ln\left|1-\sqrt{3}\right|}{\ln\left(1+\sqrt{3}\right)}\approx-0.310$,
and 
\begin{equation}
\left\{ \begin{array}{ccc}
aa & \rightarrow & aa\,ab\,ba\\
ab & \rightarrow & aa\\
ba & \rightarrow & ba\,ab\,aa
\end{array}\right.,\label{eq:seqsimple2}
\end{equation}
for which $\omega_{\text{nat}}=-1$. 

In order to obtain other similar sequences, one can start from a trial
substitution matrix for 3 letter pairs ($aa$, $ab$, and $ba$),
and calculate its largest eigenvalue, the corresponding (right) eigenvector,
and $\omega_{\text{nat}}$. The desired sequences are those with $-\infty<\omega_{\text{nat}}<0$,
yielding $\omega=0^{-}$, and having equal second and third components
of the eigenvector associated with the largest eigenvalue, which ensures
that there is no average dimerization. (The components of this eigenvector
are proportional to the fraction of the corresponding pairs in the
infinite sequence.)

\section{Perturbative renormalization group for the antiferromagnetic XXZ
chain\label{sec:apB}}

The renormalization-group (RG) equations obtained from the perturbative
approach of Vidal, Mouhanna and Giamarchi\citep{vidal99,vidal01}
are
\begin{equation}
\frac{dK}{dl}=-K^{2}\Xi\left(l\right),\label{eq:dkdl}
\end{equation}
\begin{equation}
\frac{dy_{Q}}{dl}=\left(2-K\right)y_{Q},\label{eq:dyqdl}
\end{equation}
with 
\begin{equation}
\Xi\left(l\right)=\frac{1}{2}\sum_{Q}y_{Q}^{2}\left[R\left(Q^{+}a\left(l\right)\right)+R\left(Q^{-}a\left(l\right)\right)\right],\label{eq:Xil}
\end{equation}
where $Q^{\pm}=Q\pm\pi$, the $y_{Q}=\lambda a\left|\tilde{J}\left(Q\right)\right|/u$
are initially the dimensionless Fourier components of the bonds $J_{i}$,
$\lambda$ measures the modulation strength and $l$ is a scaling
factor defined by $a\left(l\right)=a_{0}e^{l}$, the constant $a_{0}$
being proportional to the original lattice spacing. (Without loss
of generality, we take $a_{0}=1$.) $R\left(x\right)$ is a cutoff
function used to eliminate short-length degrees of freedom. We used
for $R\left(x\right)$ the precise form
\begin{equation}
R\left(x\right)=\frac{1}{1+x^{4}},
\end{equation}
but other functions having appreciable values only for $\left|x\right|<1$
yield similar results. The Luttinger parameter $K$ has an initial
value which varies with the anisotropy $\Delta$ of the XXZ chain
according to\citep{luther75,kossow2012}
\begin{equation}
K=\left[2-\frac{2}{\pi}\arccos\left(\Delta\right)\right]^{-1},
\end{equation}
therefore ranging from $K=2$ (for $\Delta=-\frac{1}{\sqrt{2}}$),
to $K=\frac{3}{2}$ (for $\Delta=-\frac{1}{2}$), to $K=1$ (for $\Delta=0$,
corresponding to the XX chain), and finally to $K=\frac{1}{2}$ (for
$\Delta=1$, corresponding to the Heisenberg chain). The correlation-length
critical exponent of the underlying dimerization transition is related
to $K$ by
\begin{equation}
\nu=\frac{1}{2-K}.\label{eq:nuK}
\end{equation}
Finally, the remaining Luttinger parameter, $u$, which appears in
the definition of $y_{Q}$, has the initial value
\begin{equation}
u=\frac{2K}{2K-1}\sin\left[\pi\left(1-\frac{1}{2K}\right)\right],
\end{equation}
corresponding to the velocity of the excitations, and its renormalization
is neglected since it only gives rise to higher order corrections\citep{vidal01}.

For a dimerized chain, in which the bonds alternate between $J_{2i}=J+\lambda/2$
and $J_{2i+1}=J-\lambda/2$, we have $\hat{J}\left(Q\right)\propto\delta\left(Q-\pi\right)$,
so we only have to worry about the renormalization of $y_{\pi}$,
whose bare value for a large chain with $N$ sites is proportional
to $N$. Starting from $K<2$, since $R\left(Q^{-}\alpha\left(l\right)\right)=R\left(0\right)=1$
for all $l$, it is clear that $K$ flows toward $0$, the strong-coupling
regime where the perturbative RG method is no longer valid. This is
consistent with the fact that dimerization opens an excitation gap
in the anisotropy regime $-\frac{1}{\sqrt{2}}<\Delta\leq1$ which
contains both the XX and the Heisenberg chains. Notice that in general
a nonzero $y_{\pi}$, even if other Fourier weights are also nonzero,
trivially leads to a runaway flow to the strong-coupling limit, with
the opening of a gap; such cases are associated with the presence
of average dimerization. In this paper we only deal with the cases
for which $y_{\pi}=0$.

Consider the case in which spins interact through nearest-neighbor
bonds $\left\{ J_{i}\right\} $ taking values $J^{\left(a\right)}$
and $J^{\left(b\right)}$ according to the sequence of letters $a$
and $b$ obtained by iterating the substitution rule in Eq.~(\ref{eq:seqomeganeg3}).
As it leads to $\omega=0^{-}$, we expect from the Harris--Luck criterion
that weak aperiodicity is irrelevant.

Indeed, it turns out that the numerical solution of the perturbative
RG equations for any finite approximant to the infinite sequence leads
to a flow in which the asymptotic value of $K$ remains close to the
initial value for all $\frac{1}{2}<K<2$, pointing to the irrelevance
of weak aperiodic modulation for the easy-plane antiferromagnetic
XXZ chain. This is related to the fact that $\tilde{J}\left(Q\right)$,
which exhibits the self-similar structure characteristic of aperiodic
sequences, has no peaks at nor in a finite neighborhood of $Q=\pi$,
as further discussed below.

\begin{figure}
\begin{centering}
\includegraphics[width=0.99\columnwidth]{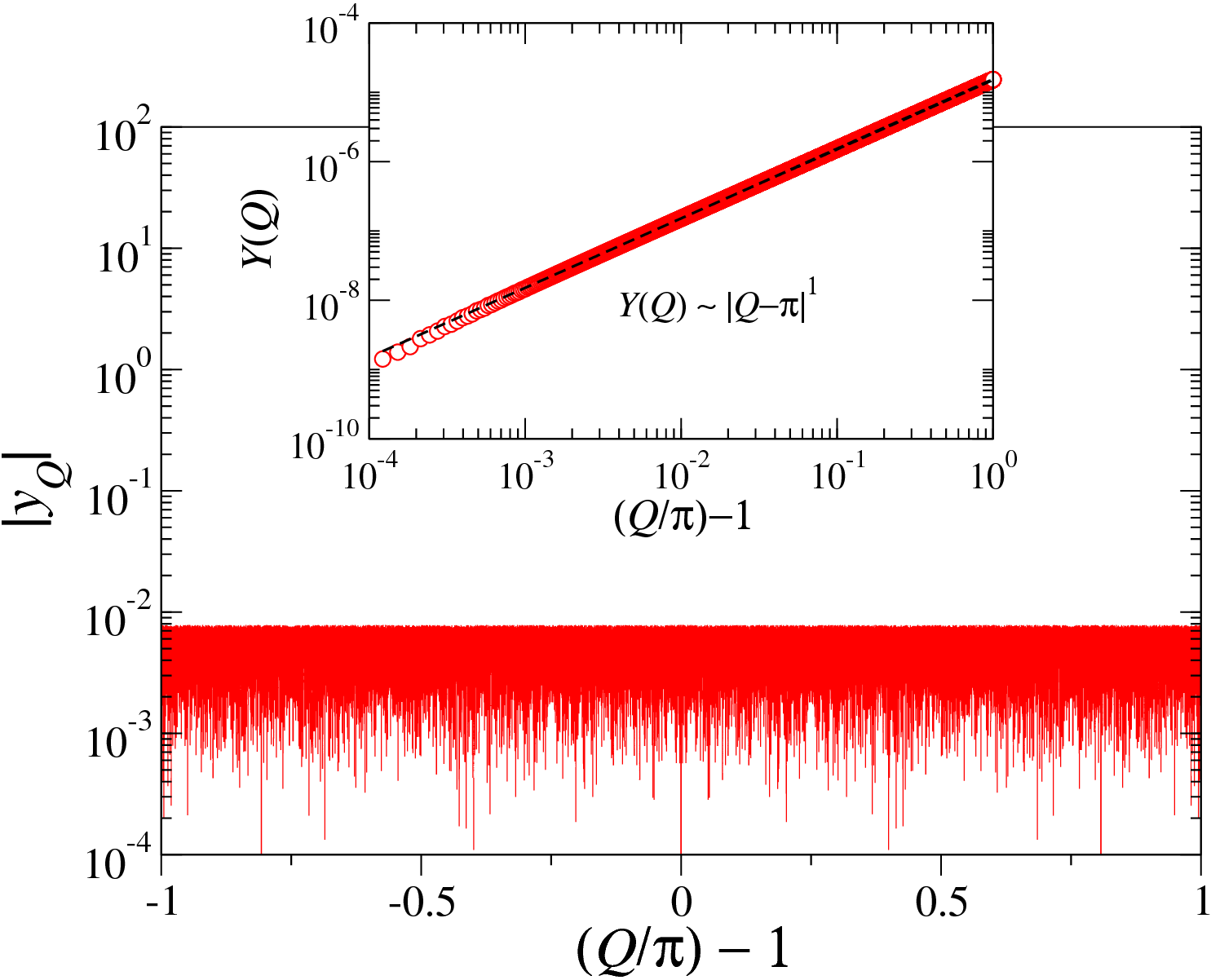}
\par\end{centering}
\caption{\label{fig:rsFFTPS}Fourier spectrum of the Rudin--Shapiro sequence.
The inset shows $Y\left(Q\right)=\sum_{q=\pi}^{Q}y^{2}\left(q\right)$,
which behaves as $\left|Q-\pi\right|^{2\alpha+1}$ if $y\left(Q\right)\sim\left|Q-\pi\right|^{\alpha}$.}
\end{figure}

\begin{figure}
\begin{centering}
\includegraphics[width=0.99\columnwidth]{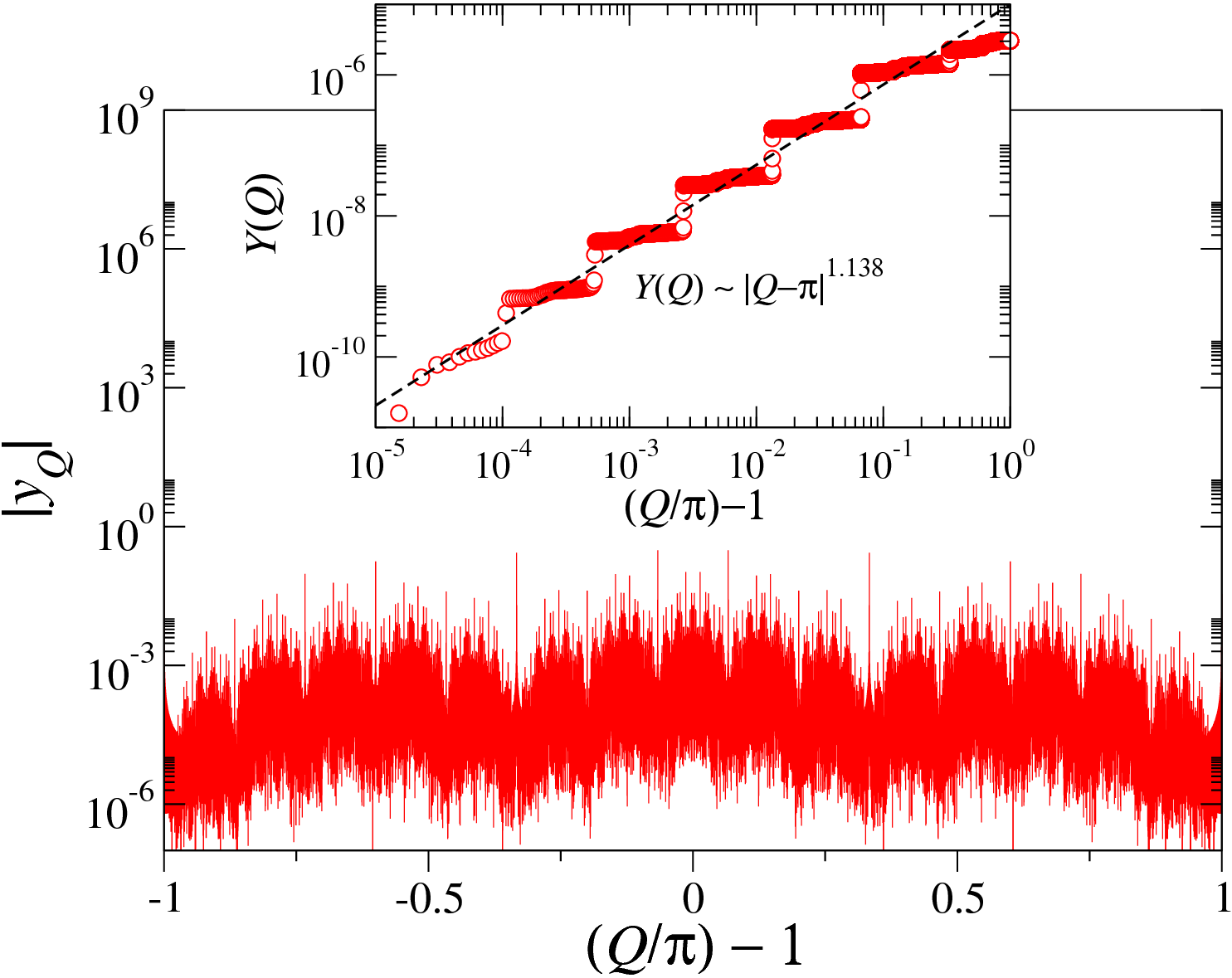}
\par\end{centering}
\caption{\label{fig:seq63FFTPS}Fourier spectrum of the 6-3 sequence. The inset
shows $Y\left(Q\right)=\sum_{q=\pi}^{Q}y^{2}\left(q\right)$, which
behaves as $\left|Q-\pi\right|^{2\alpha+1}$ if $y\left(Q\right)\sim\left|Q-\pi\right|^{\alpha}$.}
\end{figure}

In contrast, the same approach applied to the Rudin--Shapiro sequence
(\ref{eq:seqrudin}) (for which $\omega=\frac{1}{2}$), points to
its relevance in the same anisotropy regime. The same behavior is
observed for the fivefold-symmetry sequence ($\omega\approx0.285$)
and the 6-3 sequence ($\omega\approx0.431$) investigated in Ref.~\onlinecite{vieira05b}.
In all three cases, although $y_{\pi}=0$, indicating that there is
no average dimerization, the Fourier spectra are self-similar, with
peaks behaving in the neighborhood of $Q=\pi$ as $\left|Q-\pi\right|^{\alpha}$,
the constant $\alpha$ depending on the sequence, as illustrated in
Figs.\ \ref{fig:rsFFTPS} and \ref{fig:seq63FFTPS}.

This last observation allows us to attempt an approximate solution
of the perturbative RG equations. The reasoning is as follows\citep{vidal01}.
Let us assume that $K$ varies much less than $y_{Q}$ with $l$,
so that we can write
\begin{equation}
y_{Q}\left(l\right)\simeq y_{Q}\left(0\right)e^{\left(2-K\right)l},
\end{equation}
in which $y_{Q}\left(0\right)$ corresponds to the Fourier spectrum
of the original bonds. In this case, the scaling behavior of $\Xi\left(l\right)$
is given by
\begin{equation}
\Xi\left(l\right)\simeq e^{-\left(4-2K\right)l}\sum_{Q\in\mathcal{S}\left(l\right)}y_{Q}^{2}\left(0\right),
\end{equation}
where $\mathcal{S}\left(l\right)$ is the set of wavevectors, defined
by $\mathcal{S}\left(l\right)=\left\{ Q\left|\left|Q-\pi\right|\leq e^{-l}\right.\right\} $,
for which the cutoff function $R\left(Q^{-}e^{l}\right)$ is non-negligible.
Using $y_{Q}^{2}\left(0\right)\sim\left|Q-\pi\right|^{2\alpha}$,
we thus obtain
\begin{equation}
\sum_{Q\in\mathcal{S}\left(l\right)}y_{Q}^{2}\left(0\right)\sim\int_{0}^{e^{-l}}x^{2\alpha}dx\sim e^{-\left(2\alpha+1\right)l},
\end{equation}
so that
\begin{equation}
\Xi\left(l\right)\sim e^{-\left(3-2K-2\alpha\right)l}.
\end{equation}
From Eq.~(\ref{eq:dkdl}) we see that the flow of the Luttinger parameter
$K$ crucially depends on the scaling behavior of $\Xi\left(l\right)$.
If $\Xi\left(l\right)>1$ then $K$ flows to the strong-coupling limit
where the perturbative treatment breaks down, and aperiodicity is
predicted to be relevant. On the other hand, if $\Xi\left(l\right)<1$,
the flow stops at some $\lambda$-dependent finite value, and aperiodicity
is predicted to be irrelevant. For a given sequence (i.e. a given
$\alpha$), there is a critical value of $K$ separating these two
regimes:
\begin{equation}
K_{c}=\frac{3}{2}-\alpha.
\end{equation}
For $K<K_{c}$ we expect weak aperiodic modulation to be relevant.

However, Eq.~(\ref{eq:nuK}), along with the critical condition $\nu_{c}=\left(1-\omega\right)^{-1}$
derived from the Harris--Luck criterion, also implies the existence
of a critical value $K_{c}$ of the Luttinger parameter $K$, but
in terms of the wandering exponent,
\begin{equation}
K_{c}=1+\omega.\label{eq:Kc_omega}
\end{equation}
Using this last equation, derived from the Harris--Luck criterion
for the aperiodic XXZ chain, we can relate the Fourier-spectrum exponent
$\alpha$ and the pair wandering exponent $\omega$ through
\begin{equation}
\omega=\frac{1}{2}-\alpha.\label{eq:omegaalpha}
\end{equation}
The values of $\alpha$ obtained by fitting the integrated Fourier
spectra of the Rudin--Shapiro, fivefold-symmetry and 6-3 sequences
shown in Figs.~\ref{fig:rsFFTPS} and \ref{fig:seq63FFTPS} are fully
consistent with Eq.~(\ref{eq:omegaalpha}). The relation in Eq.~(\ref{eq:omegaalpha})
is also consistent with a result indicating that aperiodic fluctuations
in tight-binding Hamiltonians (equivalent to XX chains) are relevant
if, in our notation, $\alpha<\frac{1}{2}$, which corresponds to $\omega>0$;
see Ref.~\onlinecite{luck89}.

For the Fibonacci sequence, whose wandering exponent is $\omega=0^{+}$,
the relevance of weak modulation was predicted via the perturbative
renormalization-group approach in Refs.~\onlinecite{vidal99,vidal01,hida01}.
It is possible to check that the Fourier spectrum of the Fibonacci
sequence yields $\alpha=\frac{1}{2}$, again in agreement with Eq.~(\ref{eq:omegaalpha}).

\begin{figure}
\begin{centering}
\includegraphics[width=0.99\columnwidth]{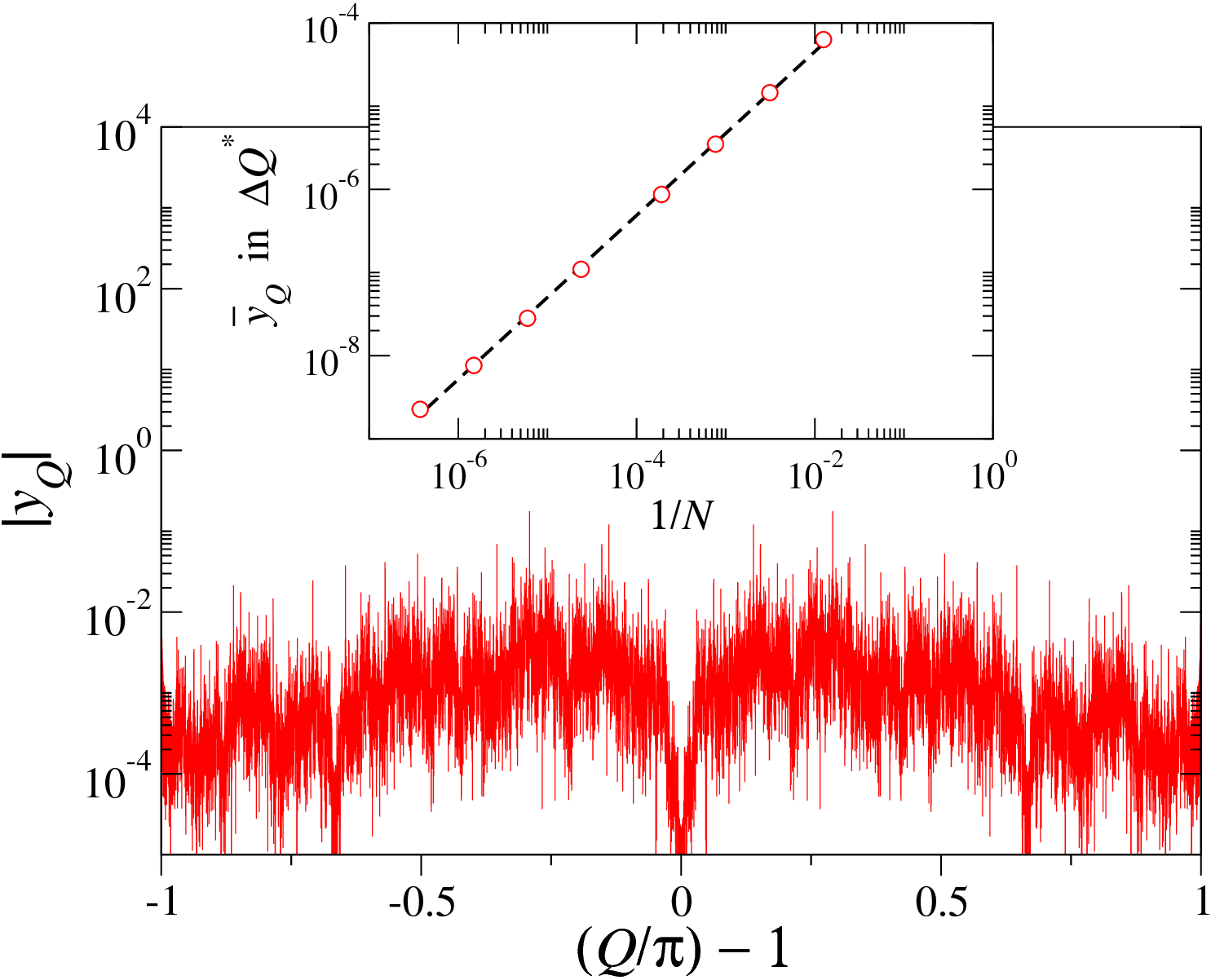}
\par\end{centering}
\caption{\label{fig:omeganegFFTPS}Fourier spectrum of the sequence in Eq.
(\ref{eq:omeganegseq}). The inset shows the finite-size scaling behavior
of the average value of the Fourier weights in a region of width $\Delta Q^{*}=10^{-3}\pi$
around $Q=\pi$.}
\end{figure}

However, the situation is different for sequences with $\omega=0^{-}$,
as those in Eq.~(\ref{eq:seqomeganeg3}) and Eqs.~(\ref{eq:seqomeganeg4})--(\ref{eq:seqsimple2}).
In this case, the Fourier spectrum has a very small and essentially
constant weight in a neighborhood of $Q=\pi$ of width $\Delta Q^{*}$
(see Fig.~\ref{fig:omeganegFFTPS}). A finite-size analysis indicates
that the weight in this region scales with the system size $N$ as
$1/N$, and thus vanishes in the thermodynamic limit $N\rightarrow\infty$.
This means that the perturbative RG flow stops at a length scale $l^{*}$
for which $e^{-l^{*}}\sim\Delta Q^{*}$. In the weak modulation limit
of $\lambda\rightarrow0$, this length scale is always reached before
any relevant flow happens, preserving the initial values of the Luttinger
parameters. Therefore, weak aperiodic bond modulation following Eqs.~(\ref{eq:seqomeganeg3})
or Eqs.~(\ref{eq:seqomeganeg4})--(\ref{eq:seqsimple2}) is \emph{irrelevant}
for the easy-plane antiferromagnetic XXZ chain, as predicted by the
Harris--Luck criterion. Moreover, Eq.~(\ref{eq:omegaalpha}) is
no longer verified, as the behavior of the Fourier spectrum around
$Q=\pi$ is not compatible with the implied value $\alpha=\frac{1}{2}$.

\section{The strong-disorder renormalization group (SDRG)\label{sec:apC}}

\begin{figure}[b]
\begin{centering}
\includegraphics[width=0.8\columnwidth]{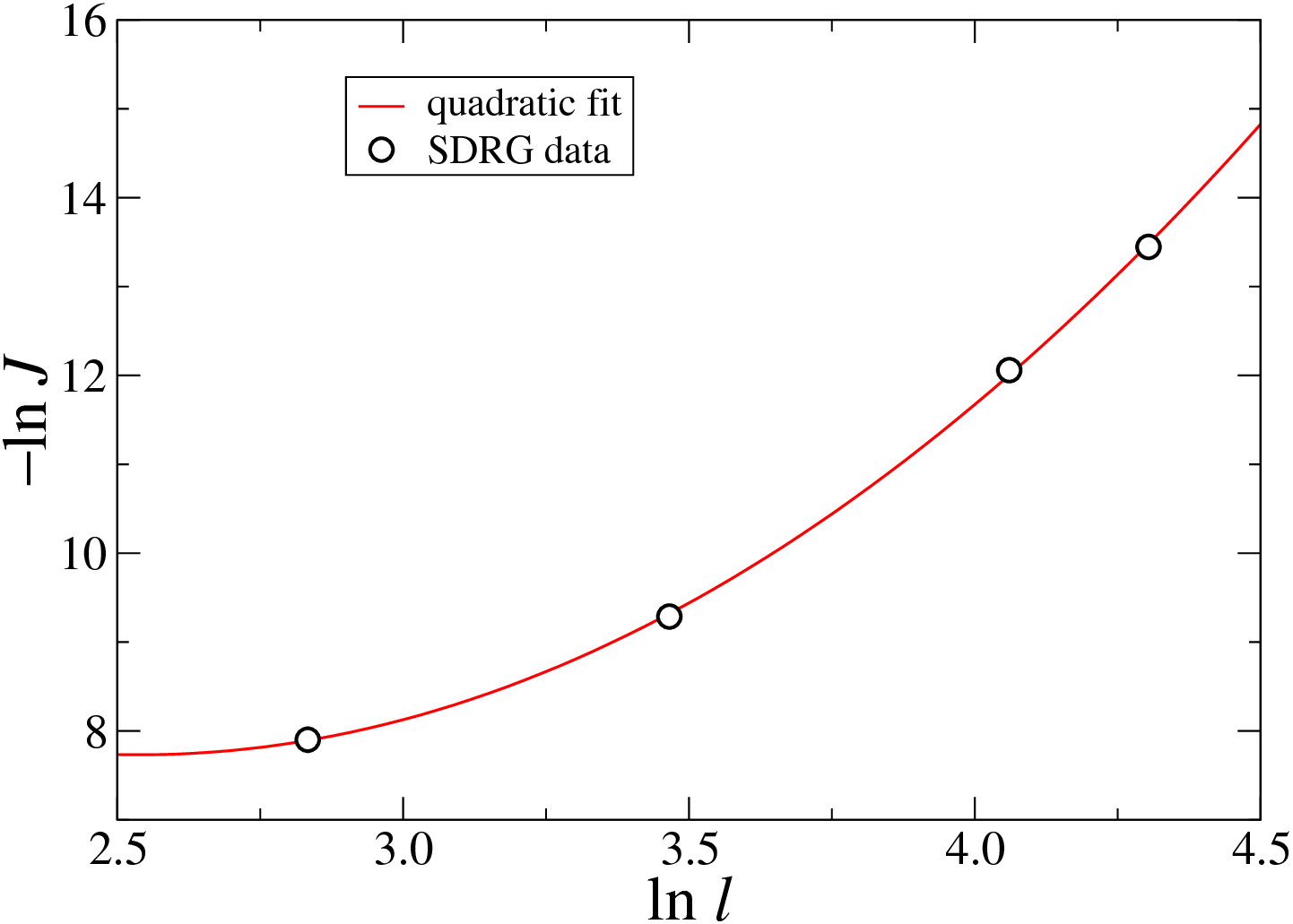}
\par\end{centering}
\caption{\label{fig:JvsL}Relation between the strengths $J$ and the lengths
$l$ of the weak effective bonds corresponding to the low-energy effective
chain when couplings follow the aperiodic sequence in Eq. (\ref{eq:seqomeganeg3}).
The continuous line is a fit using Eq. (\ref{eq:JvsL-1}). The coupling
ratio corresponds to $r=J_{a}/J_{b}=1/10$, and the strengths are
given in units of $J_{b}$. For this coupling ratio, the strength
of the strong effective bonds, as predicted by the SDRG approach,
is $\approx1.5\times10^{-3}$, with a length of 10 lattice parameters. }
\end{figure}

We consider the results of a numerical implementation of the SDRG
approach when couplings are chosen according to the sequence in Eq.~(\ref{eq:seqomeganeg3}).
As mentioned in the main text, close to the Heisenberg limit this
leads to a low-energy effective chain with emergent dimerization,
corresponding to an alternating pattern of strong and weak effective
couplings. Within the SDRG approach, the \emph{strong} effective couplings
are, suprisingly, all equal and predicted to scale as $\left(J^{\left(a\right)}\right)^{2}/J^{\left(b\right)}$.
We now investigate the relation between the magnititude $J$ and length
$l$ of the \emph{weak} effective couplings, as plotted in Fig.~\ref{fig:JvsL}
for the Heisenberg case ($\Delta=1$) and $r=0.9$. The relation can
be well fitted by
\begin{equation}
J\sim e^{-\mu\ln^{2}\left(l/l_{0}\right)},\label{eq:JvsL-1}
\end{equation}
in which $\mu$ and $l_{0}$ are constants. (Notice only four different
lengths were generated for this sequence. Other sequences may have
a different number of distinct weak effective couplings.) This form
is the same obtained for the Heisenberg chain with couplings following
the Fibonacci sequence\citep{vieira05a,vieira05b}, for which the
pair wandering exponent is $\omega=0^{+}$ but no alternating-coupling
pattern is observed. 

Similar results are also obtained from the SDRG approach for the sequences
(\ref{eq:seqomeganeg4})---(\ref{eq:seqsimple2}).

\bibliography{emergent-dimer}

\end{document}